\documentclass[%
 reprint,
 superscriptaddress,
 amsmath,amssymb,
 prl
floatfix,
]{revtex4-2}

\usepackage{graphicx}
\usepackage{dcolumn}
\usepackage{bm}

\usepackage[utf8]{inputenc}
\usepackage[T1]{fontenc}

\usepackage[english]{babel}
\usepackage{floatrow}
\usepackage[labelformat=simple, caption=false]{subfig}
\usepackage[separate-uncertainty=true]{siunitx}
\usepackage{hyperref}
\usepackage{cleveref}
\usepackage{stackengine}
\usepackage[export]{adjustbox}
\usepackage{nicefrac}

\begin{document}

\preprint{APS/123-QED}
\title{A Quantum Repeater Platform based on Single SiV$^-$ Centers in Diamond \\ with Cavity-Assisted, All-Optical Spin Access and Fast Coherent Driving}

\author{Gregor Bayer}
\thanks{These two authors contributed equally}
\author{Robert Berghaus}%
\thanks{These two authors contributed equally}
\author{Selene Sachero}
\author{Andrea B. Filipovski}
\author{Lukas Antoniuk}
\affiliation{%
 Institute for Quantum Optics, Ulm University, Albert-Einstein-Allee 11, 89081 Ulm, Germany
}%
\author{Niklas Lettner}
\affiliation{%
 Institute for Quantum Optics, Ulm University, Albert-Einstein-Allee 11, 89081 Ulm, Germany
}%
\affiliation{%
Center for Integrated Quantum Science and Technology (IQst), Ulm University,
Albert-Einstein-Allee 11, 89081 Ulm, Germany
}%
\author{Richard Waltrich}
\author{Marco Klotz}
\author{Patrick Maier}
\affiliation{%
 Institute for Quantum Optics, Ulm University, Albert-Einstein-Allee 11, 89081 Ulm, Germany
}%
\author{Viatcheslav Agafonov}
\affiliation{GREMAN, UMR 7347 CNRS, INSA-CVL, Tours University, 37200 Tours, France}

\author{Alexander Kubanek}
\thanks{Corresponding author}
\email{alexander.kubanek@uni-ulm.de}
\affiliation{%
 Institute for Quantum Optics, Ulm University, Albert-Einstein-Allee 11, 89081 Ulm, Germany
}%
\affiliation{%
Center for Integrated Quantum Science and Technology (IQst), Ulm University,
Albert-Einstein-Allee 11, 89081 Ulm, Germany
}%

\begin{abstract}
Quantum key distribution enables secure communication based on the principles of quantum mechanics. The distance in fiber-based quantum communication is limited to about a hundred kilometers due to signal attenuation. Thus, quantum repeaters are required to establish large-scale quantum networks. Ideal quantum repeater nodes possess a quantum memory which is efficiently connected to photons, the carrier of quantum information. Color centers in diamond and, in particular, the negatively-charged silicon-vacancy centers are promising candidates to establish such nodes. The major obstacle is an inefficient connection between the color centers spin to the Gaussian optics of fiber networks. Here, we present an efficient spin-photon interface. Individual silicon-vacancy centers coupled to the mode of a hemispherical Fabry-P\'erot microcavity show Purcell-factors larger than~1 when operated in a bath of liquid Helium. We demonstrate coherent optical driving with a Rabi frequency of $\SI{290}{\mega \hertz}$ and all-optical access to the electron spin in strong magnetic fields of up to $\SI{3.2}{\tesla}$. Spin initialization within $\SI{67}{\nano \second}$ with a fidelity of $\SI{80}{\percent}$ and a lifetime of $\SI{350}{\nano\second}$ are reached inside the cavity. The spin-photon interface is passively stable, enabled by placing a color center containing nanodiamond in the hemispherical Fabry-P\'erot mirror structure and by choosing short cavity lengths. Therefore, our demonstration opens the way to realize quantum repeater applications.  
  
\end{abstract}

\maketitle


\section{Introduction}
With the development of quantum computers new encryption methods become necessary to enable secure communication. Quantum key distribution (QKD) guarantees security based on the principles of quantum mechanics, where any eavesdropping attempt leaves an unavoidable fingerprint on the transmitted data \cite{pirandola_advances_2020, scarani_security_2009}. 
Photonic entanglement distribution among remote quantum nodes is the key element for quantum information exchange over long distances \cite{wehner_quantum_2018}. Fiber-based quantum channels are desirable for practical use, building on existing network technology, but are vulnerable to losses, which limits the distance for point-to-point links to about $\SI{100}{\kilo\meter}$. Currently, the most promising way to establish long-distance quantum communication, without the need for trusted nodes, is based on modular quantum repeaters \cite{briegel_quantum_1998}. Experimental realizations of quantum repeaters are extremely difficult and subject to ongoing research efforts \cite{sangouard_quantum_2011, muralidharan_optimal_2016}. The major experimental challenge is to develop elementary quantum repeater nodes with efficient fiber connectivity and, at the same time, access to long-lived quantum memories in order to reach transmission rates and state transfer fidelities that enable practical use over long distance. Recently, first demonstrations of quantum repeater nodes which contain necessary elements were demonstrated based on diamond technology \cite{bhaskar_experimental_2020}. \\
Color centers in diamond are among the most promising platforms to realize quantum repeater nodes \cite{ruf_quantum_2021,chen_building_2020,wan_large-scale_2020,bradac_quantum_2019}. Pioneering work was done with negatively-charged nitrogen-vacancy (NV$^-$) centers. A recent breakthrough marks the implementation of quantum network protocols in three-node entanglement-based quantum networks \cite{pompili_realization_2021}, where two-node entangled Bell state fidelity reaches $\SI{80}{\percent}$ with entangling rates about $\SI{9}{\hertz}$. Three-node heralded GHZ-state preparation reaches a state fidelity of $\SI{53.8}{\percent}$ and entanglement swapping across three nodes $\SI{58.7}{\percent}$ with rates of $\SI{25}{\milli\hertz}$ \cite{pompili_realization_2021}. While all
quantum repeater elements are established  \cite{rozpedek_near-term_2019,kamin_exact_2022}, scaling to multiple nodes for long distances requires significant improvements. In particular, coherent photon emission and detection requires higher efficiency in order to maintain entanglement fidelity over long-distance links at high rates. A solution are spin-photon interfaces with Purcell-enhanced coherent photon interaction. Therefore, the optical transition of the NV$^-$ center can, for example, be coupled to the mode of a Fabry-P\'erot (FP) microcavity \cite{johnson_tunable_2015,riedel_deterministic_2017,ruf_resonant_2021}. \\
Group IV color centers and, in particular, the negatively-charged silicon-vacancy (SiV$^-$) center, are an attractive alternative to NV$^-$ centers. The $\mathrm{D}_{3\mathrm{d}}$ symmetry enables spectrally stable, optical transitions and indistinguishable photons without Stark tuning \cite{rogers_multiple_2014}. High brightness and large Debye-Waller factor lead to an intrinsically higher rate of coherent photon exchange. Its electron spin coherence time reaches $\SI{13}{\milli\second}$ at $\si{\milli\kelvin}$-temperatures \cite{sukachev_silicon-vacancy_2017} and access to nuclear spins \cite{metsch_initialization_2019} potentially enables second-long memory times. But building an efficient spin-photon interface mode-matched to Gaussian optics still remains an outstanding challenge. Fast coherent optical manipulation is blocked by inefficient light-matter interaction due to the high refractive index of the diamond host. Also, the need to operate at sub-Kelvin temperatures for long spin-coherence times requires dilution refrigerators raising technical demands. \\
As a solution, we propose to work with SiV$^-$ centers in nanodiamonds (NDs) coupled to the mode of an optical FP microcavity. A locally modified phonon density of states (pDOS) could increase the operation temperature to few Kelvin, when the SiV$^-$ center is localized inside a ND \cite{klotz_prolonged_2022} making dilution refrigerators redundant. At the same time, the small size of NDs, smaller than the optical wavelength, enables integration into open resonators with low scattering loss \cite{albrecht_coupling_2013,albrecht_narrow-band_2014,johnson_tunable_2015,kaupp_purcell-enhanced_2016}. Therefore, SiV$^-$ centers in NDs with their optical transitions coupled to the mode of an optical resonator are a promising spin-photon interface for quantum repeater applications. \\
In pioneering experiments, SiV$^-$ centers in NDs \cite{benedikter_cavity-enhanced_2017} and in diamond membranes (DMs) \cite{hausler_diamond_2019} as well as germanium-vacancy (GeV$^-$) centers in DMs \cite{hoy_jensen_cavity-enhanced_2020} were coupled to the mode of fiber-based cavities at room temperature. Breakthroughs in the realization of stable FP cavities at cryogenic temperatures \cite{greuter_small_2014,janitz_fabry-perot_2015,bogdanovic_design_2017,dam_optimal_2018,janitz_cavity_2020,vadia_open-cavity_2021,fontana_mechanically_2021,ruelle_tunable_2022-1,flagan_diamond-confined_2022} enabled cavity coupling of SiV$^-$ centers at low temperatures \cite{salz_cryogenic_2020}. 
While integrated platforms for microwave control of the NV$^-$ centers electron spin inside open FP microcavities are developed \cite{bogdanovic_robust_2017}, SiV$^-$ center electron spin access is yet to be demonstrated inside open resonators. At the same time, coherent optical driving and significant Purcell-enhancement that shortens the optical lifetime is required. All these elements are essential for an efficient quantum repeater node and need to be established simultaneously. \\   
Here, we realize such a spin-photon interface by coupling individual optical transitions of single SiV$^-$ centers in a ND to the mode of an open microcavity at cryogenic temperatures. We demonstrate coherent optical driving with Rabi frequencies of $\SI{290}{\mega \hertz}$, significant lifetime shortening yielding Purcell-factors well above~1 and all-optical, cavity-assisted electronic spin access in presence of a strong magnetic field.
Therefore, we establish a promising platform to serve as a quantum repeater node which is passively stable and enables efficient mode-matching of the fundamental Gaussian $\mathrm{TEM}$-mode to fiber networks.

\section{Experimental Details}\label{sec:platform}

\begin{figure*}[htb]
    \centering 
    \includegraphics[scale=1.0]{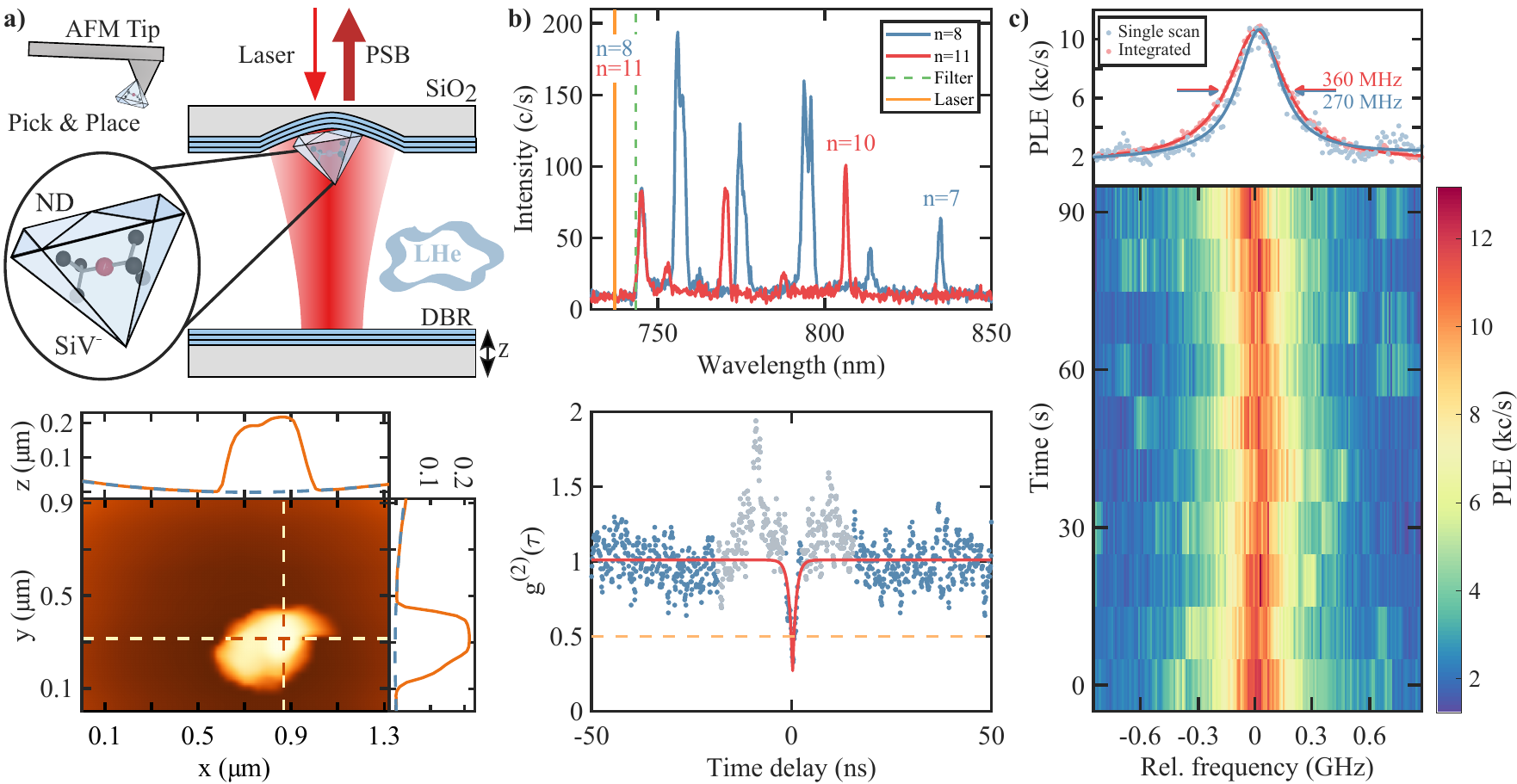}
    \caption{ \label{figure1}
    Assembling the coupled ND-microcavity system. 
    \textbf{a)} A pre-characterized ND is picked with an AFM and placed in the hemispherical Fabry-P\'erot mirror (upper panel). A subsequent AFM scan reveals the ND inside the mirror structure with a $\mathrm{RoC}$ of $\SI{8}{\micro \meter}$ indicated by blue-dashed lines (lower panel).
    \textbf{b)} Resonant PLE through the cavity mode yields SiV$^-$-sideband emission modulated by individual cavity modes shown for two different cavity lengths (upper panel). An auto-correlation measurement g$^{(2)}(\tau)$ discloses single photon emission with g$^{(2)}(0)<0.5$ (lower panel).
    \textbf{c)} The power-broadened PLE linewidth below $\SI{300}{\mega\hertz}$ (scanning speed $\approx \SI{180}{\mega\hertz\per\second}$) yields a long-term stability within $\SI{\pm12}{\mega\hertz}$ over 90 seconds. 
    }
\end{figure*}

NDs with ingrown SiV$^-$ centers are precharacterized using confocal spectroscopy, see SI. Promising NDs of small size and with spectrally stable SiV$^-$ centers are placed in the curved mirror of the FP microcavity utilizing AFM-based pick-and-place technique \cite{fehler_hybrid_2021,hausler_preparing_2019} (\cref{figure1} \textbf{a)}, upper panel). The hemispherical cavity is optimized towards small mode volume, V, and high coupling strength. Therefore, a mirror with small radius of curvature $\mathrm{RoC}\approx\SI{8}{\micro\meter}$ is fabricated using focused ion beam milling following recipes of pioneering work \cite{dolan_femtoliter_2010,kelkar_sensing_2015-1,dolan_robust_2018}. Distributed Bragg reflectors (DBRs) are coated with symmetric reflectivities on both mirrors. The transmission $\mathrm{T}=\SI{500}{ppm}$ at wavelength $\lambda=\SI{737}{\nano\meter}$ corresponds to a coating-limited Finesse of $\mathcal{F}_\mathrm{ideal}\approx6300$, see SI. After the ND-placement, the assembled ND - FP microcavity system reaches a Finesse of $\mathcal{F}=2700\pm500$ including all residual losses, such as scattering on the air-diamond interface. We optimize the system towards high passive stability by comprising moderate Finesse and a small mode volume resulting in reasonable high quality factor $Q$. Thereby, we avoid the high demands on mechanical stability and advanced locking of high-Finesse open cavities \cite{vadia_open-cavity_2021}. Furthermore, we operate the open-cavity directly in a liquid He-bath providing, in principle, infinite cooling power.

Positioning the ND in the center of the concave mirror enables highly accurate alignment of the SiV$^-$ center to the FP microcavity mode and bears the advantage of a fixed emitter-cavity coupling without the need to scan the plane mirror in $xy$-directions. Spectral overlap can be optimized in-situ by tuning the cavity length in $z$-direction. A subsequent AFM scan verifies the successful transfer of the ND to the center of the curved mirror (\cref{figure1} \textbf{a)}, lower panel). The size of the ND yields $\SI{300}{\nano\meter}$ in lateral dimensions and $\SI{200}{\nano\meter}$ in height. The SiV$^-$ center is then resonantly excited via the fundamental $\mathrm{TEM}_{00}$-mode of order $n$ with laser light at a wavelength of $\lambda_\mathrm{las}\approx\SI{736,7}{\nano\meter}$. Mode $n-1$ is located at longer wavelength and couples to the phonon sideband emission. The obtained emission spectra at $\SI{4}{\kelvin}$ are shown in \cref{figure1} \textbf{b)} (upper panel) for two different cavity lengths while keeping the same excitation wavelength $\lambda_\mathrm{las}$. In the idealized case of a planar resonator the mode wavelength is given by $\lambda_n=\frac{2}{n}\mathrm{L}$, where $\mathrm{L}$ describes the effective distance between the two mirrors. It follows that $\frac{\lambda_{n-1}}{\lambda_n}=\frac{n}{(n-1)}$. For the two cavity lengths with a difference $\Delta\mathrm{L}\approx\frac{3}{2}\lambda_\mathrm{las}$ we find $n=8$ and $n=11$ in agreement with the value for $\Delta\mathrm{L}$. Note, that the fundamental $\mathrm{TEM}_{00}$-modes with orthogonal polarization are degenerate, enabling independent control of the light polarization. \\
A second order auto-correlation measurement yields g$^{(2)}(0)=0.33\pm0.06$ without background correction, shown in \cref{figure1} \textbf{b)} (lower panel), and proves single photon emission through the cavity modes under resonant excitation. 
Note, that two peaks marked in grey arise from reflected fluorescence signal in the detectors, which is resolved in later measurements, see SI.\\
When cooled in the liquid Helium bath, narrow spectral lines become apparent in the photoluminescence excitation (PLE) scans. Individual slightly power-broadened line scans reveal linewidths $\gamma$ below $\SI{300}{\mega\hertz}$ as it would be expected for Fourier-transform (FT) limited lines for SiV$^-$ centers (\cref{figure1} \textbf{c)}). Remarkably, the cavity-assisted PLE scans disclose a long-term stability, where individual line-scans differ less than the FT-limited linewidth on timescales of $\SI{90}{\second}$. Notably, the microcavity length is not actively stabilized throughout the scans. Instead, the mechanical stability arises from the passive stability of the setup that persists even when operating in the liquid Helium bath. An estimation of the maximum length fluctuation yields that the microcavity length changes by less than one linewidth, corresponding to $\SI{160}{\pico\meter}$, see SI.

\section{Cavity-Enhanced, Coherent Optical Driving of Individual SiV\texorpdfstring{$^-$}{Lg} Centers}\label{sec:enhance}

An optical transition on resonance with the cavity mode results in a modified spontaneous decay rate. The reduction of the excited state lifetime $\tau$ is described by the Purcell-factor

\begin{equation}\label{eq:lifetime_short}
    \mathrm{F}_\mathrm{P} = \frac{\tau_\mathrm{free}}{\tau_\mathrm{cav}} = 1 + \xi f_\mathrm{P},
\end{equation}
where $\tau_\mathrm{free}$ and $\tau_\mathrm{cav}$ are the emitter lifetimes in freespace and in the cavity respectively and $\xi$ is the off-resonant branching ratio. $f_\mathrm{P}$ can be calculated as
\begin{equation}\label{eq:purcell}
    f_\mathrm{P} = \frac{3}{4\pi^2}\frac{\lambda^3}{n^3}\frac{Q}{V}
\end{equation}
with  the light wavelength $\lambda$ and the refractive index $n$ of the medium inside the microcavity. 
For our system $Q$ is primarily limited by the broader cavity resonances $ Q_{emitter} \approx  100 \times Q_{cav}$. The fundamental cavity mode volume  $V = \frac{\pi}{4}\mathrm{L} \omega_0^2$ is determined by Gaussian beam properties. With an emitter placed on the curved mirror, $V$ deviates from a linear dependence on the cavity length $\mathrm{L}$, see SI. 
Therefore, in contrast to the case of an emitter located on the flat side, $\mathrm{F}_\mathrm{P}$ strongly dependents on $\mathrm{L}$, as shown in \autoref{purcell_table} for idealized cavity parameters, suggesting higher $\mathrm{F}_\mathrm{P}$ for shorter lengths.
In a real system, additional factors such as a limited quantum efficiency, branching ratio, dipole location and orientation need to be accounted for. 
For our system, the coupling is reduced by a factor of $1.58$ as compared to an emitter located on the flat side but with aforementioned advantages for the assembled cavity system.

\begin{table}[htbp]
\caption{\label{purcell_table} Microcavity parameters for two effective lengths. From $\mathrm{RoC}$, $\mathrm{L}$ and $\mathcal{F}$, parameters such as $Q, V, \omega_0$, the ratio of beam waist on the curved to the flat side $(\nicefrac{\omega(\mathrm{L})}{\omega_0})^2$ and the respective Purcell-factors $F_\mathrm{P}^\mathrm{flat}$ and $F_\mathrm{P}^\mathrm{curv}$ are determined.}
\begin{ruledtabular}
\begin{tabular}{l|c|c|c|c|c|c|c}
    $n$ &  $\mathrm{L}_\mathrm{eff}$ $(\si{\micro\meter})$   & $Q$ & $V$ $(\lambda_\mathrm{las}^3)$ & $\omega_0$ $(\si{\micro\meter})$ & $\left(\nicefrac{\omega(\mathrm{L})}{\omega_0}\right)^2$ & $F_\mathrm{P}^\mathrm{flat}$ & $F_\mathrm{P}^\mathrm{curv}$   \\
    \hline
    8   &  $2.94$   &  22000 & 5.2 & 0.95 & 1.58 & 980 & 620   \\
    11  &  $4.05$   &  30000 & 7.5 & 0.97 & 2.03 & 950 & 470 
\end{tabular}
\end{ruledtabular}
\end{table}

\begin{figure}[htbp!]
    \centering
   \includegraphics[scale=1.0]{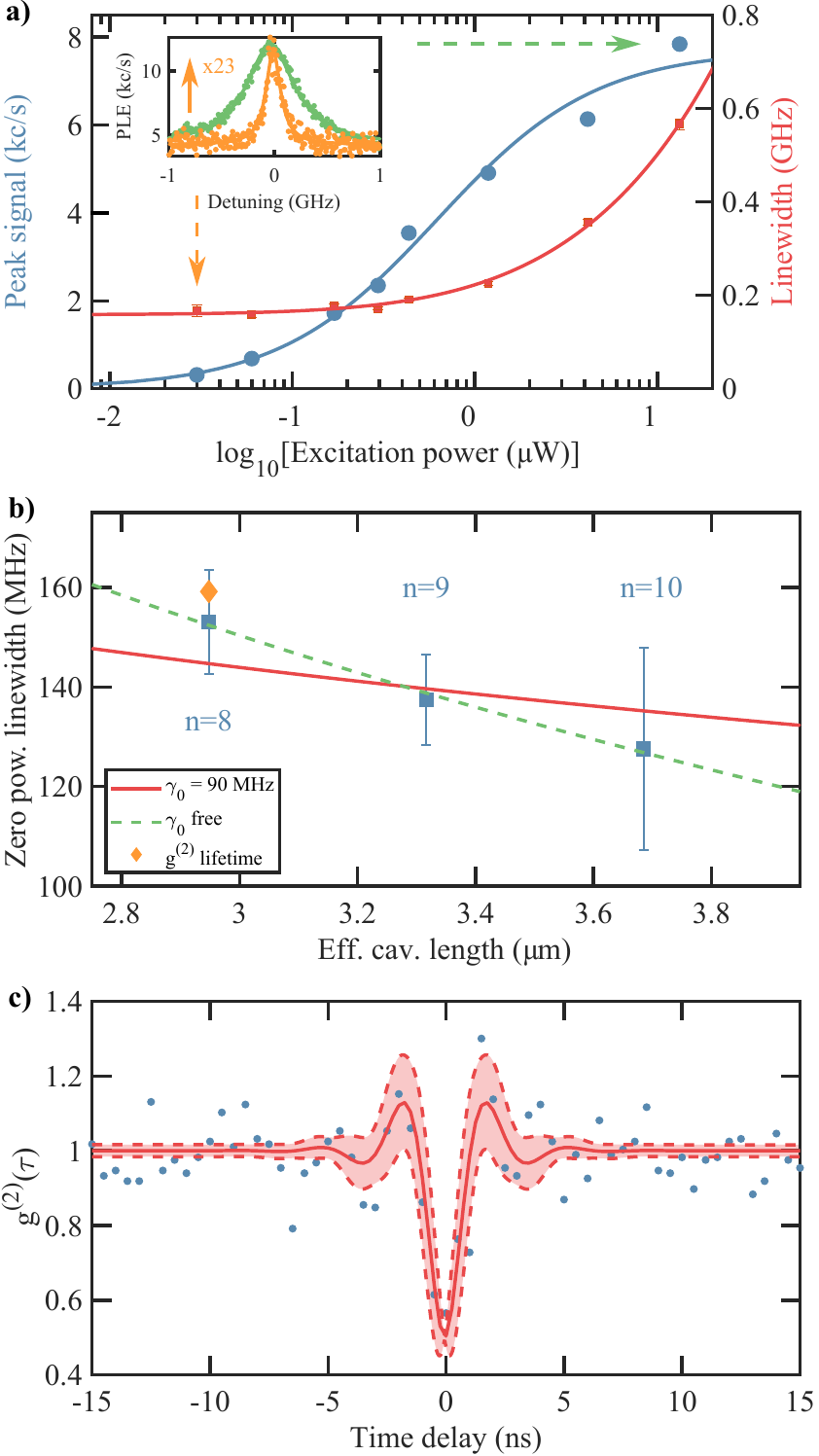}
    \caption{ \label{figure2}
    Purcell-enhancement and coherent optical driving.
    \textbf{a)} Peak count rates and linewidth from PLE scans with varying laser excitation power at the shortest accessible cavity mode $\mathrm{n}=8$. The linewidth extrapolated to zero power yields $\SI{168}{\mega\hertz}$. 
    The inset shows respective PLE scans for the highest and lowest (scaled by 23) optical power.
    \textbf{b)}     Zero-power linewidth averaged for multiple measurements at three different cavity lengths. 
    The Purcell-factor is extracted from fits with the freespace linewidth as a free parameter and fixed to $\SI{90}{\mega\hertz}$.
    \textbf{c)} Resonant $\mathrm{g}^{(2)}(\tau)$-measurement under strong drive reveal an optical lifetime of $\tau=\SI{1.0\pm0.5}{\nano\second}$ and a Rabi frequency $\Omega_\mathrm{R}/2\pi=\SI{290\pm50}{\mega\hertz}$.}%
    \end{figure}

In the experiment, we choose the mode $n=8$ and study the saturation behaviour of the emitter as shown in \cref{figure2} \textbf{a)}. Extracting to zero power yields a full width at half maximum linewidth of $\SI{168}{\mega\hertz}$. Repeating this measurement yields an averaged linewidth of $\SI{153\pm11}{\mega\hertz}$ (\cref{figure2} \textbf{b)}).
For longer cavity lengths, the averaged linewidth reduces to $\SI{128\pm21}{\mega\hertz}$ with a minimal linewidth measured of $\SI{107}{\mega\hertz}$. This suggests that the free-space emitter linewidth lies close to the FT-limit of SiV$^-$ centers in the range $90-\SI{100}{\mega\hertz}$. Assuming a FT-limited emitter linewidth, the ratio shown in \cref{eq:lifetime_short} can be reformulated as

\begin{equation}\label{eq:fwhmratio}
   \frac{\gamma_\mathrm{cav}}{\gamma_\mathrm{free}}  = \frac{\tau_\mathrm{free}}{\tau_\mathrm{cav}}. 
\end{equation}

The zero-power extracted linewidths for the SiV$^-$ center coupled at different cavity lengths is fit with two  different bounds, as shown in \cref{figure2} \textbf{b)}. A fit without any restriction on the averaged linewidth finds a free-space linewidth of $\SI{7}{\mega\hertz}$, corresponding to a lifetime $\tau_\mathrm{free}=\SI{23}{\nano\second}$ and a Purcell-factor of $21$. Usually, excited state lifetimes of 1.5-$\SI{2}{\nano\second}$ are reported \cite{rogers_multiple_2014,neu_single_2011}. Therefore, we also fit with the free-space linewidth fixed to $\SI{90}{\mega\hertz}$, corresponding to $\tau_\mathrm{free}=\SI{1.77}{\nano\second}$, and extrapolate $\mathrm{F}_\mathrm{P}=1.61\pm0.06$. The difference in the extrapolated Purcell-factors could arise from an increasing suppression of emission into the sideband as the effective cavity length shortens, as studied in detail for NV$^-$ centers in \cite{johnson_tunable_2015}. Therefore, we assume our extracted Purcell-factor of 1.61 marks a lower bound.\\
A resonant, second-order auto-correlation measurement (\cref{figure2} \textbf{c)}) with high excitation power of $\SI{7}{\micro \watt}$, corresponding to a power-broadened linewidth of $\SI{433\pm8}{\mega\hertz}$, reveals coherent optical driving of the system. The arising Rabi-oscillations are fit with a resonant g$^{(2)}(\tau)$-model, yielding an optical lifetime $\tau=\SI{1.0\pm 0.5}{\nano\second}$ and a Rabi-frequency $\Omega_\mathrm{R}/2\pi=\SI{290\pm50}{\mega\hertz}$. $\tau=\SI{1.0}{\nano\second}$ corresponds to a homogeneous linewidth of about $\SI{160}{\mega\hertz}$ (marked yellow in \cref{figure2} \textbf{b)}), in agreement with this cavity length.  \\
The ideal Purcell-factor for our cavity could reach up to 530, see \autoref{purcell_table}. Deviations from the ideal case include low quantum efficiency (order of $0.05$ \cite{chen_building_2020}) and  off-resonant branching ratio ($0.325$ \cite{zhang_strongly_2018}). An imperfect overlap of dipole and cavity field further reduces $\mathrm{F}_\mathrm{P}$.\\ 
The $\beta$-factor 
\begin{equation}
    \beta=\frac{\mathrm{F}_\mathrm{P}-1}{\mathrm{F}_\mathrm{P}} 
\end{equation}
quantifies the fraction of excited state population, that decays through coherent coupling into the microcavity mode. With the extrapolated value of $\mathrm{F}_\mathrm{P}=1.61$, we estimate $\beta \approx\SI{38}{\percent}$ as a lower bound.

\section{All-Optical Spin Access}

\begin{figure*}[htb!]
  \centering
   \includegraphics[scale=1.0]{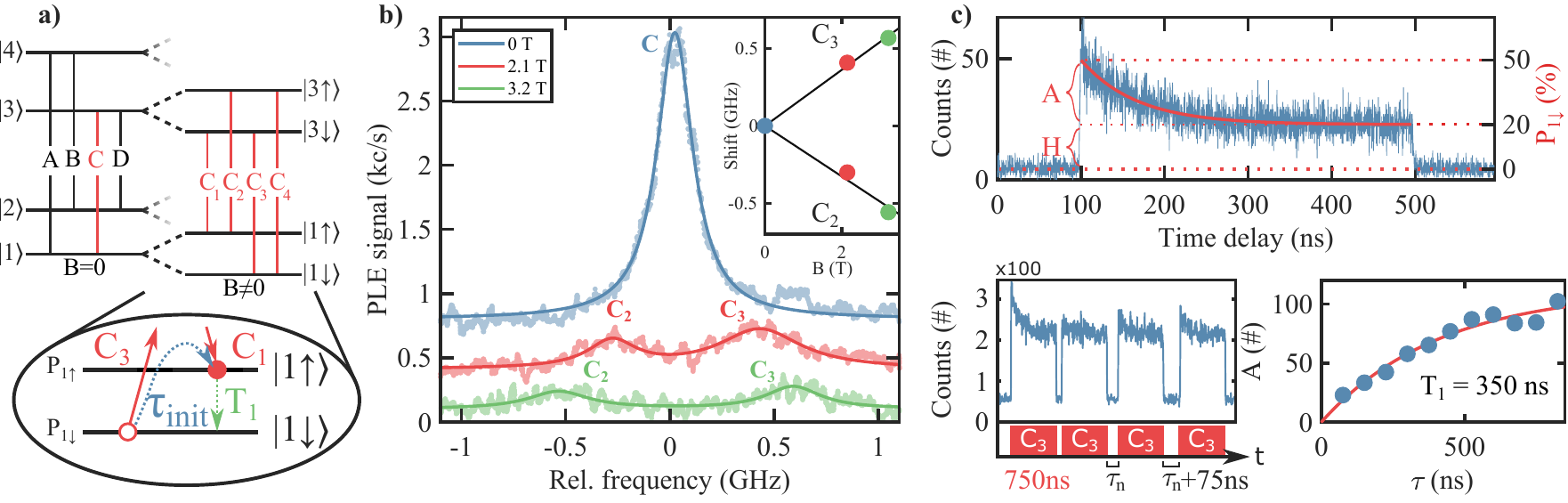}
    \caption{All-optical initialization of the SiV$^-$ centers electron spin. 
    \textbf{a)} Energy level diagram with and without external magnetic field applied along the cavity axis. In an external magnetic field the spin is initialized in the up-state by optical pumping on the spin-conserving transition $C_3$.
    \textbf{b)} The B-field dependent PLE signal shows the splitting of transition C into the spin-conserving transitions $C_2$ and $C_3$. For better visibility the scans are offset by $\SI{0.4}{\kilo c \per \second}$. 
    The inset shows the relative peak position over the field strength.  
    \textbf{c)} When pumping on resonance with $C_3$, the exponential decay of the fluorescence signal indicates the initialization in the spin-up state (upper panel). A sequence of 750ns-long laser pulses separated by an increasing waiting time by increments of $\SI{75}{\nano \second}$ yields the recovery into the equilibrium state (lower, left panel). The amplitude of an exponential fit to the decaying signal of every pulse is plotted against increasing  waiting time (lower, right panel) revealing an electron spin lifetime of $\mathrm{T}_1^{\mathrm{spin}}=\SI{350\pm40}{\nano\second}$.}%
    \label{figure3}%
\end{figure*}

We now access the electronic spin state all-optically via the cavity mode. Therefore, we split the spin-orbit states by applying a magnetic field of variable strength and resonantly address the resulting spin-cycling transitions, labeled as $C_2$ and $C_3$ in \cref{figure3} \textbf{a)}.
The cavity mode (n=8) is on resonance with the SiV$^-$ centers optical transition. PLE scans at magnetic fields of B=0, 2.1 and $\SI{3.2}{\tesla}$ reveal the field-dependent splitting into the two spin-cycling transitions $C_2$ and $C_3$, shown in \cref{figure3} \textbf{b)}. The splitting of the spin-conserving transition is a result of strain \cite{meesala_strain_2018} and magnetic field alignment to the SiV$^-$ center axis.
We attribute the reducing brightness with increasing magnetic field to a prolonged lifetime of the electron spin state as compared to the lifetime of the orbital state. Consequently, optically pumping for the duration of the spin lifetime $\mathrm{T}_1^\mathrm{spin}$ yields a reduced brightness. \\
In order to access the spin state at $\mathrm{B}=\SI{3.2}{\tesla}$, we apply $\SI{400}{\nano\second}$ long laser pulses resonant to $C_3$ at the relative frequency $\Delta \nu \approx \SI{0.6}{\giga\hertz}$, shown in \cref{figure3} \textbf{c)} (upper panel). The exponential fit to the fluorescence signal yields an initialization time $\tau_\mathrm{init}\approx\SI{67 \pm 6}{\nano\second}$, on which the system is pumped into the spin-up state. The fidelity of the electron spin initialization is extracted to $\SI{80 \pm 4}{\percent}$ by comparing the counts in the beginning and at the end of the pulse \cite{rogers_all-optical_2014}. We assume, that the system  is initially in an equal distribution of spin-up and spin-down. After optical pumping the system is preferably in spin-up state, decreasing the fluorescence signal on transition $C_3$ accordingly, see SI. 
To determine the recovery time we vary the waiting time of consecutive $\SI{750}{\nano\second}$-long pulses with an increment of $\SI{75}{\nano\second}$, as depicted in \cref{figure3} \textbf{c)} (lower, left panel). An exponential fit on the fluorescence pulse heights yields a lifetime of $\mathrm{T}_1^\mathrm{spin}=\SI{350\pm40}{\nano\second}$, see \cref{figure3} \textbf{c)} (lower, right panel). $\mathrm{T}_1^\mathrm{spin}$ is  limited by mixing of the spin states due to non-perfect magnetic field alignment, which we can not precisely control in our setup. Furthermore, remaining strain fundamentally limits the achievable $\mathrm{T}_1^\mathrm{spin}$. 
The spin lifetime, $\mathrm{T}_1^\mathrm{spin}$, and initialization, $\tau_\mathrm{init}$, are two competing processes on similar timescales in the initialization process. Therefore, we assume this to be the main limiting factor of the spin initialization fidelity. Note, that we did not observe the spin-flipping transitions, $C_1$ and $C_4$  \cite{rogers_all-optical_2014}, which we attribute to reduced fluorescence in strong magnetic fields, making the usually weaker spin-flipping transitions hardly detectable. \\
Besides the application as quantum repeater node, our platform might proof valuable as sensor at high magnetic fields. From the fit accuracy of the B-field dependent splitting in \cref{figure3} \textbf{b)} we estimate a sensitivity of $\SI{180\pm60}{\milli\tesla\hertz}^{-\nicefrac{1}{2}}$ to determine dc-field strengths in high magnetic fields, here done at $\SI{3.2}{\tesla}$. For comparison, the NV$^-$ center reach dc-magnetic field sensitivities of $\SI{40}{\nano\tesla\hertz}^{-\nicefrac{1}{2}}$, seven orders of magnitude below the sensitivity of our system \cite{rondin_magnetometry_2014}. However, magnetometry with NV$^-$ centers has limitations in high-field sensing. Typically, microwaves with frequencies on the order of $\SI{100}{\giga\hertz}$ are necessary in strong fields, requiring broadband tunable sources and compatible waveguides \cite{stepanov_high-frequency_2015}. Instead, our system requires a tunable laser, thus keeping additional technical expenses low. In the future, the sensitivity can be enhanced by employing NDs with lower strain and by improving the $\mathrm{B}$-field alignment.

\section{Conclusion and Outlook}

Summarizing, we introduce a passively-stable quantum repeater platform based on single SiV$^-$ center coupled to the mode of a FP microcavity. We coherently drive individual optical transitions through the cavity mode with a Rabi frequency of $\Omega_\mathrm{R}/2\pi=\SI{290\pm50}{\mega\hertz}$. The resulting lifetime shortening yields a Purcell-factor of $1.61$ as lower bound. We further demonstrate all-optical, electron-spin initialization with an initialization time of $\tau_\mathrm{init}\approx\SI{67}{\nano\second}$ and a fidelity of 80\%. The observed spin lifetime of $\mathrm{T}_1^\mathrm{spin}=\SI{350}{\nano\second}$ in a strong magnetic field of $\SI{3.2}{\tesla}$ is limited by field misalignment.\\
With a well-aligned magnetic field, electron spin lifetimes in the ms-range and coherence times in the $\mu$s-range become feasible at liquid Helium temperatures \cite{metsch_initialization_2019}. Furthermore, access to nuclear spin opens the possibility for longer quantum storage times with coherence times well beyond millisecond timescales \cite{sukachev_silicon-vacancy_2017,metsch_initialization_2019}. 
In addition, by further optimizing coupling parameters and reducing scattering losses higher Purcell-factors are within reach. Increasing the Finesse to $\mathcal{F}=10.000$ ($\mathcal{F}=50.000$), photon extraction efficiencies of  $\beta=\SI{69}{\percent}$ ($\beta>\SI{90}{\percent}$) and Purcell-factors above $F_\mathrm{P}=3$ ($F_\mathrm{P}=12$) are predicted for a short cavity length, see SI.
Therefore, we believe this platform will be capable of tackling the two main obstacles for quantum repeater applications, namely low operation bandwidth and rapid quantum memory dephasing.\\
Our work delivers experimental input to theoretical protocol developments \cite{van_loock_extending_2020,lee_quantum_2022} towards high secret key transmission rates over long distances exceeding hundreds of kilometers. Connecting many such quantum repeater nodes will overcome the limitations of direct quantum information transmission and lead to extended quantum networks.

\subsection{Acknowledgments}
The authors thank V.A. Davydov for synthesis and processing of the nanodiamond material. The authors thank Jason Smith for fruitful discussions. The authors gratefully acknowledge the funding by the German Federal Ministry of Education and Research (BMBF) within the project Q.Link.X (16KIS0875) and QR.X (16KISQ006). The authors gratefully acknowledge the funding by the European Union and the DFG within the Quantera-project SensExtreme. A.K. and N.L. acknowledge support of IQst. Most measurements were performed on basis of the QuDi software suite \cite{binder_qudi_2017}.

\bibliography{bibliography}

\begin{thebibliography}{52}%
\makeatletter
\providecommand \@ifxundefined [1]{%
 \@ifx{#1\undefined}
}%
\providecommand \@ifnum [1]{%
 \ifnum #1\expandafter \@firstoftwo
 \else \expandafter \@secondoftwo
 \fi
}%
\providecommand \@ifx [1]{%
 \ifx #1\expandafter \@firstoftwo
 \else \expandafter \@secondoftwo
 \fi
}%
\providecommand \natexlab [1]{#1}%
\providecommand \enquote  [1]{``#1''}%
\providecommand \bibnamefont  [1]{#1}%
\providecommand \bibfnamefont [1]{#1}%
\providecommand \citenamefont [1]{#1}%
\providecommand \href@noop [0]{\@secondoftwo}%
\providecommand \href [0]{\begingroup \@sanitize@url \@href}%
\providecommand \@href[1]{\@@startlink{#1}\@@href}%
\providecommand \@@href[1]{\endgroup#1\@@endlink}%
\providecommand \@sanitize@url [0]{\catcode `\\12\catcode `\$12\catcode
  `\&12\catcode `\#12\catcode `\^12\catcode `\_12\catcode `\%12\relax}%
\providecommand \@@startlink[1]{}%
\providecommand \@@endlink[0]{}%
\providecommand \url  [0]{\begingroup\@sanitize@url \@url }%
\providecommand \@url [1]{\endgroup\@href {#1}{\urlprefix }}%
\providecommand \urlprefix  [0]{URL }%
\providecommand \Eprint [0]{\href }%
\providecommand \doibase [0]{https://doi.org/}%
\providecommand \selectlanguage [0]{\@gobble}%
\providecommand \bibinfo  [0]{\@secondoftwo}%
\providecommand \bibfield  [0]{\@secondoftwo}%
\providecommand \translation [1]{[#1]}%
\providecommand \BibitemOpen [0]{}%
\providecommand \bibitemStop [0]{}%
\providecommand \bibitemNoStop [0]{.\EOS\space}%
\providecommand \EOS [0]{\spacefactor3000\relax}%
\providecommand \BibitemShut  [1]{\csname bibitem#1\endcsname}%
\let\auto@bib@innerbib\@empty
\bibitem [{\citenamefont {Pirandola}\ \emph {et~al.}(2020)\citenamefont
  {Pirandola}, \citenamefont {Andersen}, \citenamefont {Banchi}, \citenamefont
  {Berta}, \citenamefont {Bunandar}, \citenamefont {Colbeck}, \citenamefont
  {Englund}, \citenamefont {Gehring}, \citenamefont {Lupo}, \citenamefont
  {Ottaviani}, \citenamefont {Pereira}, \citenamefont {Razavi}, \citenamefont
  {Shamsul~Shaari}, \citenamefont {Tomamichel}, \citenamefont {Usenko},
  \citenamefont {Vallone}, \citenamefont {Villoresi},\ and\ \citenamefont
  {Wallden}}]{pirandola_advances_2020}%
  \BibitemOpen
  \bibfield  {author} {\bibinfo {author} {\bibfnamefont {S.}~\bibnamefont
  {Pirandola}}, \bibinfo {author} {\bibfnamefont {U.~L.}\ \bibnamefont
  {Andersen}}, \bibinfo {author} {\bibfnamefont {L.}~\bibnamefont {Banchi}},
  \bibinfo {author} {\bibfnamefont {M.}~\bibnamefont {Berta}}, \bibinfo
  {author} {\bibfnamefont {D.}~\bibnamefont {Bunandar}}, \bibinfo {author}
  {\bibfnamefont {R.}~\bibnamefont {Colbeck}}, \bibinfo {author} {\bibfnamefont
  {D.}~\bibnamefont {Englund}}, \bibinfo {author} {\bibfnamefont
  {T.}~\bibnamefont {Gehring}}, \bibinfo {author} {\bibfnamefont
  {C.}~\bibnamefont {Lupo}}, \bibinfo {author} {\bibfnamefont {C.}~\bibnamefont
  {Ottaviani}}, \bibinfo {author} {\bibfnamefont {J.~L.}\ \bibnamefont
  {Pereira}}, \bibinfo {author} {\bibfnamefont {M.}~\bibnamefont {Razavi}},
  \bibinfo {author} {\bibfnamefont {J.}~\bibnamefont {Shamsul~Shaari}},
  \bibinfo {author} {\bibfnamefont {M.}~\bibnamefont {Tomamichel}}, \bibinfo
  {author} {\bibfnamefont {V.~C.}\ \bibnamefont {Usenko}}, \bibinfo {author}
  {\bibfnamefont {G.}~\bibnamefont {Vallone}}, \bibinfo {author} {\bibfnamefont
  {P.}~\bibnamefont {Villoresi}},\ and\ \bibinfo {author} {\bibfnamefont
  {P.}~\bibnamefont {Wallden}},\ }\bibfield  {title} {\bibinfo {title}
  {Advances in quantum cryptography},\ }\href
  {https://doi.org/10.1364/AOP.361502} {\bibfield  {journal} {\bibinfo
  {journal} {Advances in Optics and Photonics}\ }\textbf {\bibinfo {volume}
  {12}},\ \bibinfo {pages} {1012} (\bibinfo {year} {2020})}\BibitemShut
  {NoStop}%
\bibitem [{\citenamefont {Scarani}\ \emph {et~al.}(2009)\citenamefont
  {Scarani}, \citenamefont {{Bechmann-Pasquinucci}}, \citenamefont {Cerf},
  \citenamefont {Du{\v s}ek}, \citenamefont {L{\"u}tkenhaus},\ and\
  \citenamefont {Peev}}]{scarani_security_2009}%
  \BibitemOpen
  \bibfield  {author} {\bibinfo {author} {\bibfnamefont {V.}~\bibnamefont
  {Scarani}}, \bibinfo {author} {\bibfnamefont {H.}~\bibnamefont
  {{Bechmann-Pasquinucci}}}, \bibinfo {author} {\bibfnamefont {N.~J.}\
  \bibnamefont {Cerf}}, \bibinfo {author} {\bibfnamefont {M.}~\bibnamefont
  {Du{\v s}ek}}, \bibinfo {author} {\bibfnamefont {N.}~\bibnamefont
  {L{\"u}tkenhaus}},\ and\ \bibinfo {author} {\bibfnamefont {M.}~\bibnamefont
  {Peev}},\ }\bibfield  {title} {\bibinfo {title} {The security of practical
  quantum key distribution},\ }\href
  {https://doi.org/10.1103/RevModPhys.81.1301} {\bibfield  {journal} {\bibinfo
  {journal} {Reviews of Modern Physics}\ }\textbf {\bibinfo {volume} {81}},\
  \bibinfo {pages} {1301} (\bibinfo {year} {2009})}\BibitemShut {NoStop}%
\bibitem [{\citenamefont {Wehner}\ \emph {et~al.}(2018)\citenamefont {Wehner},
  \citenamefont {Elkouss},\ and\ \citenamefont {Hanson}}]{wehner_quantum_2018}%
  \BibitemOpen
  \bibfield  {author} {\bibinfo {author} {\bibfnamefont {S.}~\bibnamefont
  {Wehner}}, \bibinfo {author} {\bibfnamefont {D.}~\bibnamefont {Elkouss}},\
  and\ \bibinfo {author} {\bibfnamefont {R.}~\bibnamefont {Hanson}},\
  }\bibfield  {title} {\bibinfo {title} {Quantum internet: {{A}} vision for the
  road ahead},\ }\href {https://doi.org/10.1126/science.aam9288} {\bibfield
  {journal} {\bibinfo  {journal} {Science}\ }\textbf {\bibinfo {volume}
  {362}},\ \bibinfo {pages} {eaam9288} (\bibinfo {year} {2018})}\BibitemShut
  {NoStop}%
\bibitem [{\citenamefont {Briegel}\ \emph {et~al.}(1998)\citenamefont
  {Briegel}, \citenamefont {D{\"u}r}, \citenamefont {Cirac},\ and\
  \citenamefont {Zoller}}]{briegel_quantum_1998}%
  \BibitemOpen
  \bibfield  {author} {\bibinfo {author} {\bibfnamefont {H.-J.}\ \bibnamefont
  {Briegel}}, \bibinfo {author} {\bibfnamefont {W.}~\bibnamefont {D{\"u}r}},
  \bibinfo {author} {\bibfnamefont {J.~I.}\ \bibnamefont {Cirac}},\ and\
  \bibinfo {author} {\bibfnamefont {P.}~\bibnamefont {Zoller}},\ }\bibfield
  {title} {\bibinfo {title} {Quantum {{Repeaters}}: {{The Role}} of {{Imperfect
  Local Operations}} in {{Quantum Communication}}},\ }\href
  {https://doi.org/10.1103/PhysRevLett.81.5932} {\bibfield  {journal} {\bibinfo
   {journal} {Physical Review Letters}\ }\textbf {\bibinfo {volume} {81}},\
  \bibinfo {pages} {5932} (\bibinfo {year} {1998})}\BibitemShut {NoStop}%
\bibitem [{\citenamefont {Sangouard}\ \emph {et~al.}(2011)\citenamefont
  {Sangouard}, \citenamefont {Simon}, \citenamefont {{de Riedmatten}},\ and\
  \citenamefont {Gisin}}]{sangouard_quantum_2011}%
  \BibitemOpen
  \bibfield  {author} {\bibinfo {author} {\bibfnamefont {N.}~\bibnamefont
  {Sangouard}}, \bibinfo {author} {\bibfnamefont {C.}~\bibnamefont {Simon}},
  \bibinfo {author} {\bibfnamefont {H.}~\bibnamefont {{de Riedmatten}}},\ and\
  \bibinfo {author} {\bibfnamefont {N.}~\bibnamefont {Gisin}},\ }\bibfield
  {title} {\bibinfo {title} {Quantum repeaters based on atomic ensembles and
  linear optics},\ }\href {https://doi.org/10.1103/RevModPhys.83.33} {\bibfield
   {journal} {\bibinfo  {journal} {Reviews of Modern Physics}\ }\textbf
  {\bibinfo {volume} {83}},\ \bibinfo {pages} {33} (\bibinfo {year}
  {2011})}\BibitemShut {NoStop}%
\bibitem [{\citenamefont {Muralidharan}\ \emph {et~al.}(2016)\citenamefont
  {Muralidharan}, \citenamefont {Li}, \citenamefont {Kim}, \citenamefont
  {L{\"u}tkenhaus}, \citenamefont {Lukin},\ and\ \citenamefont
  {Jiang}}]{muralidharan_optimal_2016}%
  \BibitemOpen
  \bibfield  {author} {\bibinfo {author} {\bibfnamefont {S.}~\bibnamefont
  {Muralidharan}}, \bibinfo {author} {\bibfnamefont {L.}~\bibnamefont {Li}},
  \bibinfo {author} {\bibfnamefont {J.}~\bibnamefont {Kim}}, \bibinfo {author}
  {\bibfnamefont {N.}~\bibnamefont {L{\"u}tkenhaus}}, \bibinfo {author}
  {\bibfnamefont {M.~D.}\ \bibnamefont {Lukin}},\ and\ \bibinfo {author}
  {\bibfnamefont {L.}~\bibnamefont {Jiang}},\ }\bibfield  {title} {\bibinfo
  {title} {Optimal architectures for long distance quantum communication},\
  }\href {https://doi.org/10.1038/srep20463} {\bibfield  {journal} {\bibinfo
  {journal} {Scientific Reports}\ }\textbf {\bibinfo {volume} {6}},\ \bibinfo
  {pages} {20463} (\bibinfo {year} {2016})}\BibitemShut {NoStop}%
\bibitem [{\citenamefont {Bhaskar}\ \emph {et~al.}(2020)\citenamefont
  {Bhaskar}, \citenamefont {Riedinger}, \citenamefont {Machielse},
  \citenamefont {Levonian}, \citenamefont {Nguyen}, \citenamefont {Knall},
  \citenamefont {Park}, \citenamefont {Englund}, \citenamefont {Lon{\v c}ar},
  \citenamefont {Sukachev},\ and\ \citenamefont
  {Lukin}}]{bhaskar_experimental_2020}%
  \BibitemOpen
  \bibfield  {author} {\bibinfo {author} {\bibfnamefont {M.~K.}\ \bibnamefont
  {Bhaskar}}, \bibinfo {author} {\bibfnamefont {R.}~\bibnamefont {Riedinger}},
  \bibinfo {author} {\bibfnamefont {B.}~\bibnamefont {Machielse}}, \bibinfo
  {author} {\bibfnamefont {D.~S.}\ \bibnamefont {Levonian}}, \bibinfo {author}
  {\bibfnamefont {C.~T.}\ \bibnamefont {Nguyen}}, \bibinfo {author}
  {\bibfnamefont {E.~N.}\ \bibnamefont {Knall}}, \bibinfo {author}
  {\bibfnamefont {H.}~\bibnamefont {Park}}, \bibinfo {author} {\bibfnamefont
  {D.}~\bibnamefont {Englund}}, \bibinfo {author} {\bibfnamefont
  {M.}~\bibnamefont {Lon{\v c}ar}}, \bibinfo {author} {\bibfnamefont {D.~D.}\
  \bibnamefont {Sukachev}},\ and\ \bibinfo {author} {\bibfnamefont {M.~D.}\
  \bibnamefont {Lukin}},\ }\bibfield  {title} {\bibinfo {title} {Experimental
  demonstration of memory-enhanced quantum communication},\ }\href
  {https://doi.org/10.1038/s41586-020-2103-5} {\bibfield  {journal} {\bibinfo
  {journal} {Nature}\ }\textbf {\bibinfo {volume} {580}},\ \bibinfo {pages}
  {60} (\bibinfo {year} {2020})}\BibitemShut {NoStop}%
\bibitem [{\citenamefont {Ruf}\ \emph {et~al.}(2021{\natexlab{a}})\citenamefont
  {Ruf}, \citenamefont {Wan}, \citenamefont {Choi}, \citenamefont {Englund},\
  and\ \citenamefont {Hanson}}]{ruf_quantum_2021}%
  \BibitemOpen
  \bibfield  {author} {\bibinfo {author} {\bibfnamefont {M.}~\bibnamefont
  {Ruf}}, \bibinfo {author} {\bibfnamefont {N.~H.}\ \bibnamefont {Wan}},
  \bibinfo {author} {\bibfnamefont {H.}~\bibnamefont {Choi}}, \bibinfo {author}
  {\bibfnamefont {D.}~\bibnamefont {Englund}},\ and\ \bibinfo {author}
  {\bibfnamefont {R.}~\bibnamefont {Hanson}},\ }\bibfield  {title} {\bibinfo
  {title} {Quantum networks based on color centers in diamond},\ }\href
  {https://doi.org/10.1063/5.0056534} {\bibfield  {journal} {\bibinfo
  {journal} {Journal of Applied Physics}\ }\textbf {\bibinfo {volume} {130}},\
  \bibinfo {pages} {070901} (\bibinfo {year} {2021}{\natexlab{a}})}\BibitemShut
  {NoStop}%
\bibitem [{\citenamefont {Chen}\ \emph {et~al.}(2020)\citenamefont {Chen},
  \citenamefont {Zheludev},\ and\ \citenamefont {Gao}}]{chen_building_2020}%
  \BibitemOpen
  \bibfield  {author} {\bibinfo {author} {\bibfnamefont {D.}~\bibnamefont
  {Chen}}, \bibinfo {author} {\bibfnamefont {N.}~\bibnamefont {Zheludev}},\
  and\ \bibinfo {author} {\bibfnamefont {W.-b.}\ \bibnamefont {Gao}},\
  }\bibfield  {title} {\bibinfo {title} {Building {{Blocks}} for {{Quantum
  Network Based}} on {{Group-IV Split-Vacancy Centers}} in {{Diamond}}},\
  }\href {https://doi.org/10.1002/qute.201900069} {\bibfield  {journal}
  {\bibinfo  {journal} {Advanced Quantum Technologies}\ }\textbf {\bibinfo
  {volume} {3}},\ \bibinfo {pages} {1900069} (\bibinfo {year}
  {2020})}\BibitemShut {NoStop}%
\bibitem [{\citenamefont {Wan}\ \emph {et~al.}(2020)\citenamefont {Wan},
  \citenamefont {Lu}, \citenamefont {Chen}, \citenamefont {Walsh},
  \citenamefont {Trusheim}, \citenamefont {De~Santis}, \citenamefont {Bersin},
  \citenamefont {Harris}, \citenamefont {Mouradian}, \citenamefont {Christen},
  \citenamefont {Bielejec},\ and\ \citenamefont
  {Englund}}]{wan_large-scale_2020}%
  \BibitemOpen
  \bibfield  {author} {\bibinfo {author} {\bibfnamefont {N.~H.}\ \bibnamefont
  {Wan}}, \bibinfo {author} {\bibfnamefont {T.-J.}\ \bibnamefont {Lu}},
  \bibinfo {author} {\bibfnamefont {K.~C.}\ \bibnamefont {Chen}}, \bibinfo
  {author} {\bibfnamefont {M.~P.}\ \bibnamefont {Walsh}}, \bibinfo {author}
  {\bibfnamefont {M.~E.}\ \bibnamefont {Trusheim}}, \bibinfo {author}
  {\bibfnamefont {L.}~\bibnamefont {De~Santis}}, \bibinfo {author}
  {\bibfnamefont {E.~A.}\ \bibnamefont {Bersin}}, \bibinfo {author}
  {\bibfnamefont {I.~B.}\ \bibnamefont {Harris}}, \bibinfo {author}
  {\bibfnamefont {S.~L.}\ \bibnamefont {Mouradian}}, \bibinfo {author}
  {\bibfnamefont {I.~R.}\ \bibnamefont {Christen}}, \bibinfo {author}
  {\bibfnamefont {E.~S.}\ \bibnamefont {Bielejec}},\ and\ \bibinfo {author}
  {\bibfnamefont {D.}~\bibnamefont {Englund}},\ }\bibfield  {title} {\bibinfo
  {title} {Large-scale integration of artificial atoms in hybrid photonic
  circuits},\ }\href {https://doi.org/10.1038/s41586-020-2441-3} {\bibfield
  {journal} {\bibinfo  {journal} {Nature}\ }\textbf {\bibinfo {volume} {583}},\
  \bibinfo {pages} {226} (\bibinfo {year} {2020})}\BibitemShut {NoStop}%
\bibitem [{\citenamefont {Bradac}\ \emph {et~al.}(2019)\citenamefont {Bradac},
  \citenamefont {Gao}, \citenamefont {Forneris}, \citenamefont {Trusheim},\
  and\ \citenamefont {Aharonovich}}]{bradac_quantum_2019}%
  \BibitemOpen
  \bibfield  {author} {\bibinfo {author} {\bibfnamefont {C.}~\bibnamefont
  {Bradac}}, \bibinfo {author} {\bibfnamefont {W.}~\bibnamefont {Gao}},
  \bibinfo {author} {\bibfnamefont {J.}~\bibnamefont {Forneris}}, \bibinfo
  {author} {\bibfnamefont {M.~E.}\ \bibnamefont {Trusheim}},\ and\ \bibinfo
  {author} {\bibfnamefont {I.}~\bibnamefont {Aharonovich}},\ }\bibfield
  {title} {\bibinfo {title} {Quantum nanophotonics with group {{IV}} defects in
  diamond},\ }\href {https://doi.org/10.1038/s41467-019-13332-w} {\bibfield
  {journal} {\bibinfo  {journal} {Nature Communications}\ }\textbf {\bibinfo
  {volume} {10}},\ \bibinfo {pages} {5625} (\bibinfo {year}
  {2019})}\BibitemShut {NoStop}%
\bibitem [{\citenamefont {Pompili}\ \emph {et~al.}(2021)\citenamefont
  {Pompili}, \citenamefont {Hermans}, \citenamefont {Baier}, \citenamefont
  {Beukers}, \citenamefont {Humphreys}, \citenamefont {Schouten}, \citenamefont
  {Vermeulen}, \citenamefont {Tiggelman}, \citenamefont {{dos Santos Martins}},
  \citenamefont {Dirkse}, \citenamefont {Wehner},\ and\ \citenamefont
  {Hanson}}]{pompili_realization_2021}%
  \BibitemOpen
  \bibfield  {author} {\bibinfo {author} {\bibfnamefont {M.}~\bibnamefont
  {Pompili}}, \bibinfo {author} {\bibfnamefont {S.~L.~N.}\ \bibnamefont
  {Hermans}}, \bibinfo {author} {\bibfnamefont {S.}~\bibnamefont {Baier}},
  \bibinfo {author} {\bibfnamefont {H.~K.~C.}\ \bibnamefont {Beukers}},
  \bibinfo {author} {\bibfnamefont {P.~C.}\ \bibnamefont {Humphreys}}, \bibinfo
  {author} {\bibfnamefont {R.~N.}\ \bibnamefont {Schouten}}, \bibinfo {author}
  {\bibfnamefont {R.~F.~L.}\ \bibnamefont {Vermeulen}}, \bibinfo {author}
  {\bibfnamefont {M.~J.}\ \bibnamefont {Tiggelman}}, \bibinfo {author}
  {\bibfnamefont {L.}~\bibnamefont {{dos Santos Martins}}}, \bibinfo {author}
  {\bibfnamefont {B.}~\bibnamefont {Dirkse}}, \bibinfo {author} {\bibfnamefont
  {S.}~\bibnamefont {Wehner}},\ and\ \bibinfo {author} {\bibfnamefont
  {R.}~\bibnamefont {Hanson}},\ }\bibfield  {title} {\bibinfo {title}
  {Realization of a multinode quantum network of remote solid-state qubits},\
  }\href {https://doi.org/10.1126/science.abg1919} {\bibfield  {journal}
  {\bibinfo  {journal} {Science}\ }\textbf {\bibinfo {volume} {372}},\ \bibinfo
  {pages} {259} (\bibinfo {year} {2021})}\BibitemShut {NoStop}%
\bibitem [{\citenamefont {Rozp{\k{e}}dek}\ \emph {et~al.}(2019)\citenamefont
  {Rozp{\k{e}}dek}, \citenamefont {Yehia}, \citenamefont {Goodenough},
  \citenamefont {Ruf}, \citenamefont {Humphreys}, \citenamefont {Hanson},
  \citenamefont {Wehner},\ and\ \citenamefont
  {Elkouss}}]{rozpedek_near-term_2019}%
  \BibitemOpen
  \bibfield  {author} {\bibinfo {author} {\bibfnamefont {F.}~\bibnamefont
  {Rozp{\k{e}}dek}}, \bibinfo {author} {\bibfnamefont {R.}~\bibnamefont
  {Yehia}}, \bibinfo {author} {\bibfnamefont {K.}~\bibnamefont {Goodenough}},
  \bibinfo {author} {\bibfnamefont {M.}~\bibnamefont {Ruf}}, \bibinfo {author}
  {\bibfnamefont {P.~C.}\ \bibnamefont {Humphreys}}, \bibinfo {author}
  {\bibfnamefont {R.}~\bibnamefont {Hanson}}, \bibinfo {author} {\bibfnamefont
  {S.}~\bibnamefont {Wehner}},\ and\ \bibinfo {author} {\bibfnamefont
  {D.}~\bibnamefont {Elkouss}},\ }\bibfield  {title} {\bibinfo {title}
  {Near-term quantum-repeater experiments with nitrogen-vacancy centers:
  {{Overcoming}} the limitations of direct transmission},\ }\href
  {https://doi.org/10.1103/PhysRevA.99.052330} {\bibfield  {journal} {\bibinfo
  {journal} {Physical Review A}\ }\textbf {\bibinfo {volume} {99}},\ \bibinfo
  {pages} {052330} (\bibinfo {year} {2019})}\BibitemShut {NoStop}%
\bibitem [{\citenamefont {Kamin}\ \emph {et~al.}(2022)\citenamefont {Kamin},
  \citenamefont {Shchukin}, \citenamefont {Schmidt},\ and\ \citenamefont {{van
  Loock}}}]{kamin_exact_2022}%
  \BibitemOpen
  \bibfield  {author} {\bibinfo {author} {\bibfnamefont {L.}~\bibnamefont
  {Kamin}}, \bibinfo {author} {\bibfnamefont {E.}~\bibnamefont {Shchukin}},
  \bibinfo {author} {\bibfnamefont {F.}~\bibnamefont {Schmidt}},\ and\ \bibinfo
  {author} {\bibfnamefont {P.}~\bibnamefont {{van Loock}}},\ }\href@noop {}
  {\bibinfo {title} {Exact rate analysis for quantum repeaters with imperfect
  memories and entanglement swapping as soon as possible}} (\bibinfo {year}
  {2022}),\ \Eprint {https://arxiv.org/abs/2203.10318} {arXiv:2203.10318
  [quant-ph]} \BibitemShut {NoStop}%
\bibitem [{\citenamefont {Johnson}\ \emph {et~al.}(2015)\citenamefont
  {Johnson}, \citenamefont {Dolan}, \citenamefont {Grange}, \citenamefont
  {Trichet}, \citenamefont {Hornecker}, \citenamefont {Chen}, \citenamefont
  {Weng}, \citenamefont {Hughes}, \citenamefont {Watt}, \citenamefont
  {Auff{\`e}ves},\ and\ \citenamefont {Smith}}]{johnson_tunable_2015}%
  \BibitemOpen
  \bibfield  {author} {\bibinfo {author} {\bibfnamefont {S.}~\bibnamefont
  {Johnson}}, \bibinfo {author} {\bibfnamefont {P.~R.}\ \bibnamefont {Dolan}},
  \bibinfo {author} {\bibfnamefont {T.}~\bibnamefont {Grange}}, \bibinfo
  {author} {\bibfnamefont {A.~A.~P.}\ \bibnamefont {Trichet}}, \bibinfo
  {author} {\bibfnamefont {G.}~\bibnamefont {Hornecker}}, \bibinfo {author}
  {\bibfnamefont {Y.~C.}\ \bibnamefont {Chen}}, \bibinfo {author}
  {\bibfnamefont {L.}~\bibnamefont {Weng}}, \bibinfo {author} {\bibfnamefont
  {G.~M.}\ \bibnamefont {Hughes}}, \bibinfo {author} {\bibfnamefont {A.~A.~R.}\
  \bibnamefont {Watt}}, \bibinfo {author} {\bibfnamefont {A.}~\bibnamefont
  {Auff{\`e}ves}},\ and\ \bibinfo {author} {\bibfnamefont {J.~M.}\ \bibnamefont
  {Smith}},\ }\bibfield  {title} {\bibinfo {title} {Tunable cavity coupling of
  the zero phonon line of a nitrogen-vacancy defect in diamond},\ }\href
  {https://doi.org/10.1088/1367-2630/17/12/122003} {\bibfield  {journal}
  {\bibinfo  {journal} {New Journal of Physics}\ }\textbf {\bibinfo {volume}
  {17}},\ \bibinfo {pages} {122003} (\bibinfo {year} {2015})}\BibitemShut
  {NoStop}%
\bibitem [{\citenamefont {Riedel}\ \emph {et~al.}(2017)\citenamefont {Riedel},
  \citenamefont {S{\"o}llner}, \citenamefont {Shields}, \citenamefont
  {Starosielec}, \citenamefont {Appel}, \citenamefont {Neu}, \citenamefont
  {Maletinsky},\ and\ \citenamefont {Warburton}}]{riedel_deterministic_2017}%
  \BibitemOpen
  \bibfield  {author} {\bibinfo {author} {\bibfnamefont {D.}~\bibnamefont
  {Riedel}}, \bibinfo {author} {\bibfnamefont {I.}~\bibnamefont {S{\"o}llner}},
  \bibinfo {author} {\bibfnamefont {B.~J.}\ \bibnamefont {Shields}}, \bibinfo
  {author} {\bibfnamefont {S.}~\bibnamefont {Starosielec}}, \bibinfo {author}
  {\bibfnamefont {P.}~\bibnamefont {Appel}}, \bibinfo {author} {\bibfnamefont
  {E.}~\bibnamefont {Neu}}, \bibinfo {author} {\bibfnamefont {P.}~\bibnamefont
  {Maletinsky}},\ and\ \bibinfo {author} {\bibfnamefont {R.~J.}\ \bibnamefont
  {Warburton}},\ }\bibfield  {title} {\bibinfo {title} {Deterministic
  {{Enhancement}} of {{Coherent Photon Generation}} from a {{Nitrogen-Vacancy
  Center}} in {{Ultrapure Diamond}}},\ }\href
  {https://doi.org/10.1103/PhysRevX.7.031040} {\bibfield  {journal} {\bibinfo
  {journal} {Physical Review X}\ }\textbf {\bibinfo {volume} {7}},\ \bibinfo
  {pages} {031040} (\bibinfo {year} {2017})}\BibitemShut {NoStop}%
\bibitem [{\citenamefont {Ruf}\ \emph {et~al.}(2021{\natexlab{b}})\citenamefont
  {Ruf}, \citenamefont {Weaver}, \citenamefont {{van Dam}},\ and\ \citenamefont
  {Hanson}}]{ruf_resonant_2021}%
  \BibitemOpen
  \bibfield  {author} {\bibinfo {author} {\bibfnamefont {M.}~\bibnamefont
  {Ruf}}, \bibinfo {author} {\bibfnamefont {M.}~\bibnamefont {Weaver}},
  \bibinfo {author} {\bibfnamefont {S.}~\bibnamefont {{van Dam}}},\ and\
  \bibinfo {author} {\bibfnamefont {R.}~\bibnamefont {Hanson}},\ }\bibfield
  {title} {\bibinfo {title} {Resonant {{Excitation}} and {{Purcell
  Enhancement}} of {{Coherent Nitrogen-Vacancy Centers Coupled}} to a
  {{Fabry-Perot Microcavity}}},\ }\href
  {https://doi.org/10.1103/PhysRevApplied.15.024049} {\bibfield  {journal}
  {\bibinfo  {journal} {Physical Review Applied}\ }\textbf {\bibinfo {volume}
  {15}},\ \bibinfo {pages} {024049} (\bibinfo {year}
  {2021}{\natexlab{b}})}\BibitemShut {NoStop}%
\bibitem [{\citenamefont {Rogers}\ \emph
  {et~al.}(2014{\natexlab{a}})\citenamefont {Rogers}, \citenamefont {Jahnke},
  \citenamefont {Teraji}, \citenamefont {Marseglia}, \citenamefont
  {M{\"u}ller}, \citenamefont {Naydenov}, \citenamefont {Schauffert},
  \citenamefont {Kranz}, \citenamefont {Isoya}, \citenamefont {McGuinness},\
  and\ \citenamefont {Jelezko}}]{rogers_multiple_2014}%
  \BibitemOpen
  \bibfield  {author} {\bibinfo {author} {\bibfnamefont {L.}~\bibnamefont
  {Rogers}}, \bibinfo {author} {\bibfnamefont {K.}~\bibnamefont {Jahnke}},
  \bibinfo {author} {\bibfnamefont {T.}~\bibnamefont {Teraji}}, \bibinfo
  {author} {\bibfnamefont {L.}~\bibnamefont {Marseglia}}, \bibinfo {author}
  {\bibfnamefont {C.}~\bibnamefont {M{\"u}ller}}, \bibinfo {author}
  {\bibfnamefont {B.}~\bibnamefont {Naydenov}}, \bibinfo {author}
  {\bibfnamefont {H.}~\bibnamefont {Schauffert}}, \bibinfo {author}
  {\bibfnamefont {C.}~\bibnamefont {Kranz}}, \bibinfo {author} {\bibfnamefont
  {J.}~\bibnamefont {Isoya}}, \bibinfo {author} {\bibfnamefont
  {L.}~\bibnamefont {McGuinness}},\ and\ \bibinfo {author} {\bibfnamefont
  {F.}~\bibnamefont {Jelezko}},\ }\bibfield  {title} {\bibinfo {title}
  {Multiple intrinsically identical single-photon emitters in the solid
  state},\ }\href {https://doi.org/10.1038/ncomms5739} {\bibfield  {journal}
  {\bibinfo  {journal} {Nature Communications}\ }\textbf {\bibinfo {volume}
  {5}},\ \bibinfo {pages} {4739} (\bibinfo {year}
  {2014}{\natexlab{a}})}\BibitemShut {NoStop}%
\bibitem [{\citenamefont {Sukachev}\ \emph {et~al.}(2017)\citenamefont
  {Sukachev}, \citenamefont {Sipahigil}, \citenamefont {Nguyen}, \citenamefont
  {Bhaskar}, \citenamefont {Evans}, \citenamefont {Jelezko},\ and\
  \citenamefont {Lukin}}]{sukachev_silicon-vacancy_2017}%
  \BibitemOpen
  \bibfield  {author} {\bibinfo {author} {\bibfnamefont {D.~D.}\ \bibnamefont
  {Sukachev}}, \bibinfo {author} {\bibfnamefont {A.}~\bibnamefont {Sipahigil}},
  \bibinfo {author} {\bibfnamefont {C.~T.}\ \bibnamefont {Nguyen}}, \bibinfo
  {author} {\bibfnamefont {M.~K.}\ \bibnamefont {Bhaskar}}, \bibinfo {author}
  {\bibfnamefont {R.~E.}\ \bibnamefont {Evans}}, \bibinfo {author}
  {\bibfnamefont {F.}~\bibnamefont {Jelezko}},\ and\ \bibinfo {author}
  {\bibfnamefont {M.~D.}\ \bibnamefont {Lukin}},\ }\bibfield  {title} {\bibinfo
  {title} {Silicon-{{Vacancy Spin Qubit}} in {{Diamond}}: {{A Quantum Memory
  Exceeding}} 10 ms with {{Single-Shot State Readout}}},\ }\href
  {https://doi.org/10.1103/PhysRevLett.119.223602} {\bibfield  {journal}
  {\bibinfo  {journal} {Physical Review Letters}\ }\textbf {\bibinfo {volume}
  {119}},\ \bibinfo {pages} {223602} (\bibinfo {year} {2017})}\BibitemShut
  {NoStop}%
\bibitem [{\citenamefont {Metsch}\ \emph {et~al.}(2019)\citenamefont {Metsch},
  \citenamefont {Senkalla}, \citenamefont {Tratzmiller}, \citenamefont
  {Scheuer}, \citenamefont {Kern}, \citenamefont {Achard}, \citenamefont
  {Tallaire}, \citenamefont {Plenio}, \citenamefont {Siyushev},\ and\
  \citenamefont {Jelezko}}]{metsch_initialization_2019}%
  \BibitemOpen
  \bibfield  {author} {\bibinfo {author} {\bibfnamefont {M.~H.}\ \bibnamefont
  {Metsch}}, \bibinfo {author} {\bibfnamefont {K.}~\bibnamefont {Senkalla}},
  \bibinfo {author} {\bibfnamefont {B.}~\bibnamefont {Tratzmiller}}, \bibinfo
  {author} {\bibfnamefont {J.}~\bibnamefont {Scheuer}}, \bibinfo {author}
  {\bibfnamefont {M.}~\bibnamefont {Kern}}, \bibinfo {author} {\bibfnamefont
  {J.}~\bibnamefont {Achard}}, \bibinfo {author} {\bibfnamefont
  {A.}~\bibnamefont {Tallaire}}, \bibinfo {author} {\bibfnamefont {M.~B.}\
  \bibnamefont {Plenio}}, \bibinfo {author} {\bibfnamefont {P.}~\bibnamefont
  {Siyushev}},\ and\ \bibinfo {author} {\bibfnamefont {F.}~\bibnamefont
  {Jelezko}},\ }\bibfield  {title} {\bibinfo {title} {Initialization and
  {{Readout}} of {{Nuclear Spins}} via a {{Negatively Charged Silicon-Vacancy
  Center}} in {{Diamond}}},\ }\href
  {https://doi.org/10.1103/PhysRevLett.122.190503} {\bibfield  {journal}
  {\bibinfo  {journal} {Physical Review Letters}\ }\textbf {\bibinfo {volume}
  {122}},\ \bibinfo {pages} {190503} (\bibinfo {year} {2019})}\BibitemShut
  {NoStop}%
\bibitem [{\citenamefont {Klotz}\ \emph {et~al.}(2022)\citenamefont {Klotz},
  \citenamefont {Fehler}, \citenamefont {Waltrich}, \citenamefont {Steiger},
  \citenamefont {H{\"a}u{\ss}ler}, \citenamefont {Reddy}, \citenamefont
  {Kulikova}, \citenamefont {Davydov}, \citenamefont {Agafonov}, \citenamefont
  {Doherty},\ and\ \citenamefont {Kubanek}}]{klotz_prolonged_2022}%
  \BibitemOpen
  \bibfield  {author} {\bibinfo {author} {\bibfnamefont {M.}~\bibnamefont
  {Klotz}}, \bibinfo {author} {\bibfnamefont {K.~G.}\ \bibnamefont {Fehler}},
  \bibinfo {author} {\bibfnamefont {R.}~\bibnamefont {Waltrich}}, \bibinfo
  {author} {\bibfnamefont {E.~S.}\ \bibnamefont {Steiger}}, \bibinfo {author}
  {\bibfnamefont {S.}~\bibnamefont {H{\"a}u{\ss}ler}}, \bibinfo {author}
  {\bibfnamefont {P.}~\bibnamefont {Reddy}}, \bibinfo {author} {\bibfnamefont
  {L.~F.}\ \bibnamefont {Kulikova}}, \bibinfo {author} {\bibfnamefont {V.~A.}\
  \bibnamefont {Davydov}}, \bibinfo {author} {\bibfnamefont {V.~N.}\
  \bibnamefont {Agafonov}}, \bibinfo {author} {\bibfnamefont {M.~W.}\
  \bibnamefont {Doherty}},\ and\ \bibinfo {author} {\bibfnamefont
  {A.}~\bibnamefont {Kubanek}},\ }\bibfield  {title} {\bibinfo {title}
  {Prolonged {{Orbital Relaxation}} by {{Locally Modified Phonon Density}} of
  {{States}} for the {{SiV- Center}} in {{Nanodiamonds}}},\ }\href
  {https://doi.org/10.1103/PhysRevLett.128.153602} {\bibfield  {journal}
  {\bibinfo  {journal} {Physical Review Letters}\ }\textbf {\bibinfo {volume}
  {128}},\ \bibinfo {pages} {153602} (\bibinfo {year} {2022})}\BibitemShut
  {NoStop}%
\bibitem [{\citenamefont {Albrecht}\ \emph {et~al.}(2013)\citenamefont
  {Albrecht}, \citenamefont {Bommer}, \citenamefont {Deutsch}, \citenamefont
  {Reichel},\ and\ \citenamefont {Becher}}]{albrecht_coupling_2013}%
  \BibitemOpen
  \bibfield  {author} {\bibinfo {author} {\bibfnamefont {R.}~\bibnamefont
  {Albrecht}}, \bibinfo {author} {\bibfnamefont {A.}~\bibnamefont {Bommer}},
  \bibinfo {author} {\bibfnamefont {C.}~\bibnamefont {Deutsch}}, \bibinfo
  {author} {\bibfnamefont {J.}~\bibnamefont {Reichel}},\ and\ \bibinfo {author}
  {\bibfnamefont {C.}~\bibnamefont {Becher}},\ }\bibfield  {title} {\bibinfo
  {title} {Coupling of a {{Single Nitrogen-Vacancy Center}} in {{Diamond}} to a
  {{Fiber-Based Microcavity}}},\ }\href
  {https://doi.org/10.1103/PhysRevLett.110.243602} {\bibfield  {journal}
  {\bibinfo  {journal} {Physical Review Letters}\ }\textbf {\bibinfo {volume}
  {110}},\ \bibinfo {pages} {243602} (\bibinfo {year} {2013})}\BibitemShut
  {NoStop}%
\bibitem [{\citenamefont {Albrecht}\ \emph {et~al.}(2014)\citenamefont
  {Albrecht}, \citenamefont {Bommer}, \citenamefont {Pauly}, \citenamefont
  {M{\"u}cklich}, \citenamefont {Schell}, \citenamefont {Engel}, \citenamefont
  {Schr{\"o}der}, \citenamefont {Benson}, \citenamefont {Reichel},\ and\
  \citenamefont {Becher}}]{albrecht_narrow-band_2014}%
  \BibitemOpen
  \bibfield  {author} {\bibinfo {author} {\bibfnamefont {R.}~\bibnamefont
  {Albrecht}}, \bibinfo {author} {\bibfnamefont {A.}~\bibnamefont {Bommer}},
  \bibinfo {author} {\bibfnamefont {C.}~\bibnamefont {Pauly}}, \bibinfo
  {author} {\bibfnamefont {F.}~\bibnamefont {M{\"u}cklich}}, \bibinfo {author}
  {\bibfnamefont {A.~W.}\ \bibnamefont {Schell}}, \bibinfo {author}
  {\bibfnamefont {P.}~\bibnamefont {Engel}}, \bibinfo {author} {\bibfnamefont
  {T.}~\bibnamefont {Schr{\"o}der}}, \bibinfo {author} {\bibfnamefont
  {O.}~\bibnamefont {Benson}}, \bibinfo {author} {\bibfnamefont
  {J.}~\bibnamefont {Reichel}},\ and\ \bibinfo {author} {\bibfnamefont
  {C.}~\bibnamefont {Becher}},\ }\bibfield  {title} {\bibinfo {title}
  {Narrow-band single photon emission at room temperature based on a single
  nitrogen-vacancy center coupled to an all-fiber-cavity},\ }\href
  {https://doi.org/10.1063/1.4893612} {\bibfield  {journal} {\bibinfo
  {journal} {Applied Physics Letters}\ }\textbf {\bibinfo {volume} {105}},\
  \bibinfo {pages} {073113} (\bibinfo {year} {2014})}\BibitemShut {NoStop}%
\bibitem [{\citenamefont {Kaupp}\ \emph {et~al.}(2016)\citenamefont {Kaupp},
  \citenamefont {H{\"u}mmer}, \citenamefont {Mader}, \citenamefont
  {Schlederer}, \citenamefont {Benedikter}, \citenamefont {Haeusser},
  \citenamefont {Chang}, \citenamefont {Fedder}, \citenamefont {H{\"a}nsch},\
  and\ \citenamefont {Hunger}}]{kaupp_purcell-enhanced_2016}%
  \BibitemOpen
  \bibfield  {author} {\bibinfo {author} {\bibfnamefont {H.}~\bibnamefont
  {Kaupp}}, \bibinfo {author} {\bibfnamefont {T.}~\bibnamefont {H{\"u}mmer}},
  \bibinfo {author} {\bibfnamefont {M.}~\bibnamefont {Mader}}, \bibinfo
  {author} {\bibfnamefont {B.}~\bibnamefont {Schlederer}}, \bibinfo {author}
  {\bibfnamefont {J.}~\bibnamefont {Benedikter}}, \bibinfo {author}
  {\bibfnamefont {P.}~\bibnamefont {Haeusser}}, \bibinfo {author}
  {\bibfnamefont {H.-C.}\ \bibnamefont {Chang}}, \bibinfo {author}
  {\bibfnamefont {H.}~\bibnamefont {Fedder}}, \bibinfo {author} {\bibfnamefont
  {T.~W.}\ \bibnamefont {H{\"a}nsch}},\ and\ \bibinfo {author} {\bibfnamefont
  {D.}~\bibnamefont {Hunger}},\ }\bibfield  {title} {\bibinfo {title}
  {Purcell-{{Enhanced Single-Photon Emission}} from {{Nitrogen-Vacancy Centers
  Coupled}} to a {{Tunable Microcavity}}},\ }\href
  {https://doi.org/10.1103/PhysRevApplied.6.054010} {\bibfield  {journal}
  {\bibinfo  {journal} {Physical Review Applied}\ }\textbf {\bibinfo {volume}
  {6}},\ \bibinfo {pages} {054010} (\bibinfo {year} {2016})}\BibitemShut
  {NoStop}%
\bibitem [{\citenamefont {Benedikter}\ \emph {et~al.}(2017)\citenamefont
  {Benedikter}, \citenamefont {Kaupp}, \citenamefont {H{\"u}mmer},
  \citenamefont {Liang}, \citenamefont {Bommer}, \citenamefont {Becher},
  \citenamefont {Krueger}, \citenamefont {Smith}, \citenamefont {H{\"a}nsch},\
  and\ \citenamefont {Hunger}}]{benedikter_cavity-enhanced_2017}%
  \BibitemOpen
  \bibfield  {author} {\bibinfo {author} {\bibfnamefont {J.}~\bibnamefont
  {Benedikter}}, \bibinfo {author} {\bibfnamefont {H.}~\bibnamefont {Kaupp}},
  \bibinfo {author} {\bibfnamefont {T.}~\bibnamefont {H{\"u}mmer}}, \bibinfo
  {author} {\bibfnamefont {Y.}~\bibnamefont {Liang}}, \bibinfo {author}
  {\bibfnamefont {A.}~\bibnamefont {Bommer}}, \bibinfo {author} {\bibfnamefont
  {C.}~\bibnamefont {Becher}}, \bibinfo {author} {\bibfnamefont
  {A.}~\bibnamefont {Krueger}}, \bibinfo {author} {\bibfnamefont {J.~M.}\
  \bibnamefont {Smith}}, \bibinfo {author} {\bibfnamefont {T.~W.}\ \bibnamefont
  {H{\"a}nsch}},\ and\ \bibinfo {author} {\bibfnamefont {D.}~\bibnamefont
  {Hunger}},\ }\bibfield  {title} {\bibinfo {title} {Cavity-{{Enhanced
  Single-Photon Source Based}} on the {{Silicon-Vacancy Center}} in
  {{Diamond}}},\ }\href {https://doi.org/10.1103/PhysRevApplied.7.024031}
  {\bibfield  {journal} {\bibinfo  {journal} {Physical Review Applied}\
  }\textbf {\bibinfo {volume} {7}},\ \bibinfo {pages} {024031} (\bibinfo {year}
  {2017})}\BibitemShut {NoStop}%
\bibitem [{\citenamefont {H{\"a}u{\ss}ler}\ \emph
  {et~al.}(2019{\natexlab{a}})\citenamefont {H{\"a}u{\ss}ler}, \citenamefont
  {Benedikter}, \citenamefont {Bray}, \citenamefont {Regan}, \citenamefont
  {Dietrich}, \citenamefont {Twamley}, \citenamefont {Aharonovich},
  \citenamefont {Hunger},\ and\ \citenamefont
  {Kubanek}}]{hausler_diamond_2019}%
  \BibitemOpen
  \bibfield  {author} {\bibinfo {author} {\bibfnamefont {S.}~\bibnamefont
  {H{\"a}u{\ss}ler}}, \bibinfo {author} {\bibfnamefont {J.}~\bibnamefont
  {Benedikter}}, \bibinfo {author} {\bibfnamefont {K.}~\bibnamefont {Bray}},
  \bibinfo {author} {\bibfnamefont {B.}~\bibnamefont {Regan}}, \bibinfo
  {author} {\bibfnamefont {A.}~\bibnamefont {Dietrich}}, \bibinfo {author}
  {\bibfnamefont {J.}~\bibnamefont {Twamley}}, \bibinfo {author} {\bibfnamefont
  {I.}~\bibnamefont {Aharonovich}}, \bibinfo {author} {\bibfnamefont
  {D.}~\bibnamefont {Hunger}},\ and\ \bibinfo {author} {\bibfnamefont
  {A.}~\bibnamefont {Kubanek}},\ }\bibfield  {title} {\bibinfo {title} {Diamond
  photonics platform based on silicon vacancy centers in a single-crystal
  diamond membrane and a fiber cavity},\ }\href
  {https://doi.org/10.1103/PhysRevB.99.165310} {\bibfield  {journal} {\bibinfo
  {journal} {Physical Review B}\ }\textbf {\bibinfo {volume} {99}},\ \bibinfo
  {pages} {165310} (\bibinfo {year} {2019}{\natexlab{a}})}\BibitemShut
  {NoStop}%
\bibitem [{\citenamefont {H{\o}y~Jensen}\ \emph {et~al.}(2020)\citenamefont
  {H{\o}y~Jensen}, \citenamefont {Janitz}, \citenamefont {Fontana},
  \citenamefont {He}, \citenamefont {Gobron}, \citenamefont {Radko},
  \citenamefont {Bhaskar}, \citenamefont {Evans}, \citenamefont
  {Rodr{\'i}guez~Rosenblueth}, \citenamefont {Childress}, \citenamefont
  {Huck},\ and\ \citenamefont
  {Lund~Andersen}}]{hoy_jensen_cavity-enhanced_2020}%
  \BibitemOpen
  \bibfield  {author} {\bibinfo {author} {\bibfnamefont {R.}~\bibnamefont
  {H{\o}y~Jensen}}, \bibinfo {author} {\bibfnamefont {E.}~\bibnamefont
  {Janitz}}, \bibinfo {author} {\bibfnamefont {Y.}~\bibnamefont {Fontana}},
  \bibinfo {author} {\bibfnamefont {Y.}~\bibnamefont {He}}, \bibinfo {author}
  {\bibfnamefont {O.}~\bibnamefont {Gobron}}, \bibinfo {author} {\bibfnamefont
  {I.~P.}\ \bibnamefont {Radko}}, \bibinfo {author} {\bibfnamefont
  {M.}~\bibnamefont {Bhaskar}}, \bibinfo {author} {\bibfnamefont
  {R.}~\bibnamefont {Evans}}, \bibinfo {author} {\bibfnamefont {C.~D.}\
  \bibnamefont {Rodr{\'i}guez~Rosenblueth}}, \bibinfo {author} {\bibfnamefont
  {L.}~\bibnamefont {Childress}}, \bibinfo {author} {\bibfnamefont
  {A.}~\bibnamefont {Huck}},\ and\ \bibinfo {author} {\bibfnamefont
  {U.}~\bibnamefont {Lund~Andersen}},\ }\bibfield  {title} {\bibinfo {title}
  {Cavity-{{Enhanced Photon Emission}} from a {{Single Germanium-Vacancy
  Center}} in a {{Diamond Membrane}}},\ }\href
  {https://doi.org/10.1103/PhysRevApplied.13.064016} {\bibfield  {journal}
  {\bibinfo  {journal} {Physical Review Applied}\ }\textbf {\bibinfo {volume}
  {13}},\ \bibinfo {pages} {064016} (\bibinfo {year} {2020})}\BibitemShut
  {NoStop}%
\bibitem [{\citenamefont {Greuter}\ \emph {et~al.}(2014)\citenamefont
  {Greuter}, \citenamefont {Starosielec}, \citenamefont {Najer}, \citenamefont
  {Ludwig}, \citenamefont {Duempelmann}, \citenamefont {Rohner},\ and\
  \citenamefont {Warburton}}]{greuter_small_2014}%
  \BibitemOpen
  \bibfield  {author} {\bibinfo {author} {\bibfnamefont {L.}~\bibnamefont
  {Greuter}}, \bibinfo {author} {\bibfnamefont {S.}~\bibnamefont
  {Starosielec}}, \bibinfo {author} {\bibfnamefont {D.}~\bibnamefont {Najer}},
  \bibinfo {author} {\bibfnamefont {A.}~\bibnamefont {Ludwig}}, \bibinfo
  {author} {\bibfnamefont {L.}~\bibnamefont {Duempelmann}}, \bibinfo {author}
  {\bibfnamefont {D.}~\bibnamefont {Rohner}},\ and\ \bibinfo {author}
  {\bibfnamefont {R.~J.}\ \bibnamefont {Warburton}},\ }\bibfield  {title}
  {\bibinfo {title} {A small mode volume tunable microcavity: {{Development}}
  and characterization},\ }\href {https://doi.org/10.1063/1.4896415} {\bibfield
   {journal} {\bibinfo  {journal} {Applied Physics Letters}\ }\textbf {\bibinfo
  {volume} {105}},\ \bibinfo {pages} {121105} (\bibinfo {year}
  {2014})}\BibitemShut {NoStop}%
\bibitem [{\citenamefont {Janitz}\ \emph {et~al.}(2015)\citenamefont {Janitz},
  \citenamefont {Ruf}, \citenamefont {Dimock}, \citenamefont {Bourassa},
  \citenamefont {Sankey},\ and\ \citenamefont
  {Childress}}]{janitz_fabry-perot_2015}%
  \BibitemOpen
  \bibfield  {author} {\bibinfo {author} {\bibfnamefont {E.}~\bibnamefont
  {Janitz}}, \bibinfo {author} {\bibfnamefont {M.}~\bibnamefont {Ruf}},
  \bibinfo {author} {\bibfnamefont {M.}~\bibnamefont {Dimock}}, \bibinfo
  {author} {\bibfnamefont {A.}~\bibnamefont {Bourassa}}, \bibinfo {author}
  {\bibfnamefont {J.}~\bibnamefont {Sankey}},\ and\ \bibinfo {author}
  {\bibfnamefont {L.}~\bibnamefont {Childress}},\ }\bibfield  {title} {\bibinfo
  {title} {Fabry-{{Perot}} microcavity for diamond-based photonics},\ }\href
  {https://doi.org/10.1103/PhysRevA.92.043844} {\bibfield  {journal} {\bibinfo
  {journal} {Physical Review A}\ }\textbf {\bibinfo {volume} {92}},\ \bibinfo
  {pages} {043844} (\bibinfo {year} {2015})}\BibitemShut {NoStop}%
\bibitem [{\citenamefont {Bogdanovi{\'c}}\ \emph
  {et~al.}(2017{\natexlab{a}})\citenamefont {Bogdanovi{\'c}}, \citenamefont
  {{van Dam}}, \citenamefont {Bonato}, \citenamefont {Coenen}, \citenamefont
  {Zwerver}, \citenamefont {Hensen}, \citenamefont {Liddy}, \citenamefont
  {Fink}, \citenamefont {Reiserer}, \citenamefont {Lon{\v c}ar},\ and\
  \citenamefont {Hanson}}]{bogdanovic_design_2017}%
  \BibitemOpen
  \bibfield  {author} {\bibinfo {author} {\bibfnamefont {S.}~\bibnamefont
  {Bogdanovi{\'c}}}, \bibinfo {author} {\bibfnamefont {S.~B.}\ \bibnamefont
  {{van Dam}}}, \bibinfo {author} {\bibfnamefont {C.}~\bibnamefont {Bonato}},
  \bibinfo {author} {\bibfnamefont {L.~C.}\ \bibnamefont {Coenen}}, \bibinfo
  {author} {\bibfnamefont {A.-M.~J.}\ \bibnamefont {Zwerver}}, \bibinfo
  {author} {\bibfnamefont {B.}~\bibnamefont {Hensen}}, \bibinfo {author}
  {\bibfnamefont {M.~S.~Z.}\ \bibnamefont {Liddy}}, \bibinfo {author}
  {\bibfnamefont {T.}~\bibnamefont {Fink}}, \bibinfo {author} {\bibfnamefont
  {A.}~\bibnamefont {Reiserer}}, \bibinfo {author} {\bibfnamefont
  {M.}~\bibnamefont {Lon{\v c}ar}},\ and\ \bibinfo {author} {\bibfnamefont
  {R.}~\bibnamefont {Hanson}},\ }\bibfield  {title} {\bibinfo {title} {Design
  and low-temperature characterization of a tunable microcavity for
  diamond-based quantum networks},\ }\href {https://doi.org/10.1063/1.4982168}
  {\bibfield  {journal} {\bibinfo  {journal} {Applied Physics Letters}\
  }\textbf {\bibinfo {volume} {110}},\ \bibinfo {pages} {171103} (\bibinfo
  {year} {2017}{\natexlab{a}})}\BibitemShut {NoStop}%
\bibitem [{\citenamefont {van Dam}\ \emph {et~al.}(2018)\citenamefont {van
  Dam}, \citenamefont {Ruf},\ and\ \citenamefont {Hanson}}]{dam_optimal_2018}%
  \BibitemOpen
  \bibfield  {author} {\bibinfo {author} {\bibfnamefont {S.~B.}\ \bibnamefont
  {van Dam}}, \bibinfo {author} {\bibfnamefont {M.}~\bibnamefont {Ruf}},\ and\
  \bibinfo {author} {\bibfnamefont {R.}~\bibnamefont {Hanson}},\ }\bibfield
  {title} {\bibinfo {title} {Optimal design of diamond-air microcavities for
  quantum networks using an analytical approach},\ }\href
  {https://doi.org/10.1088/1367-2630/aaec29} {\bibfield  {journal} {\bibinfo
  {journal} {New Journal of Physics}\ }\textbf {\bibinfo {volume} {20}},\
  \bibinfo {pages} {115004} (\bibinfo {year} {2018})}\BibitemShut {NoStop}%
\bibitem [{\citenamefont {Janitz}\ \emph {et~al.}(2020)\citenamefont {Janitz},
  \citenamefont {Bhaskar},\ and\ \citenamefont
  {Childress}}]{janitz_cavity_2020}%
  \BibitemOpen
  \bibfield  {author} {\bibinfo {author} {\bibfnamefont {E.}~\bibnamefont
  {Janitz}}, \bibinfo {author} {\bibfnamefont {M.~K.}\ \bibnamefont
  {Bhaskar}},\ and\ \bibinfo {author} {\bibfnamefont {L.}~\bibnamefont
  {Childress}},\ }\bibfield  {title} {\bibinfo {title} {Cavity quantum
  electrodynamics with color centers in diamond},\ }\href
  {https://doi.org/10.1364/OPTICA.398628} {\bibfield  {journal} {\bibinfo
  {journal} {Optica}\ }\textbf {\bibinfo {volume} {7}},\ \bibinfo {pages}
  {1232} (\bibinfo {year} {2020})}\BibitemShut {NoStop}%
\bibitem [{\citenamefont {Vadia}\ \emph {et~al.}(2021)\citenamefont {Vadia},
  \citenamefont {Scherzer}, \citenamefont {Thierschmann}, \citenamefont
  {Sch{\"a}fermeier}, \citenamefont {Dal~Savio}, \citenamefont {Taniguchi},
  \citenamefont {Watanabe}, \citenamefont {Hunger}, \citenamefont
  {Karra{\"i}},\ and\ \citenamefont {H{\"o}gele}}]{vadia_open-cavity_2021}%
  \BibitemOpen
  \bibfield  {author} {\bibinfo {author} {\bibfnamefont {S.}~\bibnamefont
  {Vadia}}, \bibinfo {author} {\bibfnamefont {J.}~\bibnamefont {Scherzer}},
  \bibinfo {author} {\bibfnamefont {H.}~\bibnamefont {Thierschmann}}, \bibinfo
  {author} {\bibfnamefont {C.}~\bibnamefont {Sch{\"a}fermeier}}, \bibinfo
  {author} {\bibfnamefont {C.}~\bibnamefont {Dal~Savio}}, \bibinfo {author}
  {\bibfnamefont {T.}~\bibnamefont {Taniguchi}}, \bibinfo {author}
  {\bibfnamefont {K.}~\bibnamefont {Watanabe}}, \bibinfo {author}
  {\bibfnamefont {D.}~\bibnamefont {Hunger}}, \bibinfo {author} {\bibfnamefont
  {K.}~\bibnamefont {Karra{\"i}}},\ and\ \bibinfo {author} {\bibfnamefont
  {A.}~\bibnamefont {H{\"o}gele}},\ }\bibfield  {title} {\bibinfo {title}
  {Open-{{Cavity}} in {{Closed-Cycle Cryostat}} as a {{Quantum Optics
  Platform}}},\ }\href {https://doi.org/10.1103/PRXQuantum.2.040318} {\bibfield
   {journal} {\bibinfo  {journal} {PRX Quantum}\ }\textbf {\bibinfo {volume}
  {2}},\ \bibinfo {pages} {040318} (\bibinfo {year} {2021})}\BibitemShut
  {NoStop}%
\bibitem [{\citenamefont {Fontana}\ \emph {et~al.}(2021)\citenamefont
  {Fontana}, \citenamefont {Zifkin}, \citenamefont {Janitz}, \citenamefont
  {Rodr{\'i}guez~Rosenblueth},\ and\ \citenamefont
  {Childress}}]{fontana_mechanically_2021}%
  \BibitemOpen
  \bibfield  {author} {\bibinfo {author} {\bibfnamefont {Y.}~\bibnamefont
  {Fontana}}, \bibinfo {author} {\bibfnamefont {R.}~\bibnamefont {Zifkin}},
  \bibinfo {author} {\bibfnamefont {E.}~\bibnamefont {Janitz}}, \bibinfo
  {author} {\bibfnamefont {C.~D.}\ \bibnamefont {Rodr{\'i}guez~Rosenblueth}},\
  and\ \bibinfo {author} {\bibfnamefont {L.}~\bibnamefont {Childress}},\
  }\bibfield  {title} {\bibinfo {title} {A mechanically stable and tunable
  cryogenic {{Fabry}}\textendash{{P\'erot}} microcavity},\ }\href
  {https://doi.org/10.1063/5.0049520} {\bibfield  {journal} {\bibinfo
  {journal} {Review of Scientific Instruments}\ }\textbf {\bibinfo {volume}
  {92}},\ \bibinfo {pages} {053906} (\bibinfo {year} {2021})}\BibitemShut
  {NoStop}%
\bibitem [{\citenamefont {Ruelle}\ \emph {et~al.}(2022)\citenamefont {Ruelle},
  \citenamefont {Jaeger}, \citenamefont {Fogliano}, \citenamefont {Braakman},\
  and\ \citenamefont {Poggio}}]{ruelle_tunable_2022-1}%
  \BibitemOpen
  \bibfield  {author} {\bibinfo {author} {\bibfnamefont {T.}~\bibnamefont
  {Ruelle}}, \bibinfo {author} {\bibfnamefont {D.}~\bibnamefont {Jaeger}},
  \bibinfo {author} {\bibfnamefont {F.}~\bibnamefont {Fogliano}}, \bibinfo
  {author} {\bibfnamefont {F.}~\bibnamefont {Braakman}},\ and\ \bibinfo
  {author} {\bibfnamefont {M.}~\bibnamefont {Poggio}},\ }\bibfield  {title}
  {\bibinfo {title} {A tunable fiber {{Fabry}}\textendash{{Perot}} cavity for
  hybrid optomechanics stabilized at 4 {{K}}},\ }\href
  {https://doi.org/10.1063/5.0098140} {\bibfield  {journal} {\bibinfo
  {journal} {Review of Scientific Instruments}\ }\textbf {\bibinfo {volume}
  {93}},\ \bibinfo {pages} {095003} (\bibinfo {year} {2022})}\BibitemShut
  {NoStop}%
\bibitem [{\citenamefont {Fl{\aa}gan}\ \emph {et~al.}(2022)\citenamefont
  {Fl{\aa}gan}, \citenamefont {Riedel}, \citenamefont {Javadi}, \citenamefont
  {Jakubczyk}, \citenamefont {Maletinsky},\ and\ \citenamefont
  {Warburton}}]{flagan_diamond-confined_2022}%
  \BibitemOpen
  \bibfield  {author} {\bibinfo {author} {\bibfnamefont {S.}~\bibnamefont
  {Fl{\aa}gan}}, \bibinfo {author} {\bibfnamefont {D.}~\bibnamefont {Riedel}},
  \bibinfo {author} {\bibfnamefont {A.}~\bibnamefont {Javadi}}, \bibinfo
  {author} {\bibfnamefont {T.}~\bibnamefont {Jakubczyk}}, \bibinfo {author}
  {\bibfnamefont {P.}~\bibnamefont {Maletinsky}},\ and\ \bibinfo {author}
  {\bibfnamefont {R.~J.}\ \bibnamefont {Warburton}},\ }\bibfield  {title}
  {\bibinfo {title} {A diamond-confined open microcavity featuring a high
  quality-factor and a small mode-volume},\ }\href
  {https://doi.org/10.1063/5.0081577} {\bibfield  {journal} {\bibinfo
  {journal} {Journal of Applied Physics}\ }\textbf {\bibinfo {volume} {131}},\
  \bibinfo {pages} {113102} (\bibinfo {year} {2022})}\BibitemShut {NoStop}%
\bibitem [{\citenamefont {Salz}\ \emph {et~al.}(2020)\citenamefont {Salz},
  \citenamefont {Herrmann}, \citenamefont {Nadarajah}, \citenamefont {Stahl},
  \citenamefont {Hettrich}, \citenamefont {Stacey}, \citenamefont {Prawer},
  \citenamefont {Hunger},\ and\ \citenamefont
  {{Schmidt-Kaler}}}]{salz_cryogenic_2020}%
  \BibitemOpen
  \bibfield  {author} {\bibinfo {author} {\bibfnamefont {M.}~\bibnamefont
  {Salz}}, \bibinfo {author} {\bibfnamefont {Y.}~\bibnamefont {Herrmann}},
  \bibinfo {author} {\bibfnamefont {A.}~\bibnamefont {Nadarajah}}, \bibinfo
  {author} {\bibfnamefont {A.}~\bibnamefont {Stahl}}, \bibinfo {author}
  {\bibfnamefont {M.}~\bibnamefont {Hettrich}}, \bibinfo {author}
  {\bibfnamefont {A.}~\bibnamefont {Stacey}}, \bibinfo {author} {\bibfnamefont
  {S.}~\bibnamefont {Prawer}}, \bibinfo {author} {\bibfnamefont
  {D.}~\bibnamefont {Hunger}},\ and\ \bibinfo {author} {\bibfnamefont
  {F.}~\bibnamefont {{Schmidt-Kaler}}},\ }\bibfield  {title} {\bibinfo {title}
  {Cryogenic platform for coupling color centers in diamond membranes to a
  fiber-based microcavity},\ }\href
  {https://doi.org/10.1007/s00340-020-07478-5} {\bibfield  {journal} {\bibinfo
  {journal} {Applied Physics B}\ }\textbf {\bibinfo {volume} {126}},\ \bibinfo
  {pages} {131} (\bibinfo {year} {2020})}\BibitemShut {NoStop}%
\bibitem [{\citenamefont {Bogdanovi{\'c}}\ \emph
  {et~al.}(2017{\natexlab{b}})\citenamefont {Bogdanovi{\'c}}, \citenamefont
  {Liddy}, \citenamefont {{van Dam}}, \citenamefont {Coenen}, \citenamefont
  {Fink}, \citenamefont {Lon{\v c}ar},\ and\ \citenamefont
  {Hanson}}]{bogdanovic_robust_2017}%
  \BibitemOpen
  \bibfield  {author} {\bibinfo {author} {\bibfnamefont {S.}~\bibnamefont
  {Bogdanovi{\'c}}}, \bibinfo {author} {\bibfnamefont {M.~S.~Z.}\ \bibnamefont
  {Liddy}}, \bibinfo {author} {\bibfnamefont {S.~B.}\ \bibnamefont {{van
  Dam}}}, \bibinfo {author} {\bibfnamefont {L.~C.}\ \bibnamefont {Coenen}},
  \bibinfo {author} {\bibfnamefont {T.}~\bibnamefont {Fink}}, \bibinfo {author}
  {\bibfnamefont {M.}~\bibnamefont {Lon{\v c}ar}},\ and\ \bibinfo {author}
  {\bibfnamefont {R.}~\bibnamefont {Hanson}},\ }\bibfield  {title} {\bibinfo
  {title} {Robust nano-fabrication of an integrated platform for spin control
  in a tunable microcavity},\ }\href {https://doi.org/10.1063/1.5001144}
  {\bibfield  {journal} {\bibinfo  {journal} {APL Photonics}\ }\textbf
  {\bibinfo {volume} {2}},\ \bibinfo {pages} {126101} (\bibinfo {year}
  {2017}{\natexlab{b}})}\BibitemShut {NoStop}%
\bibitem [{\citenamefont {Fehler}\ \emph {et~al.}(2021)\citenamefont {Fehler},
  \citenamefont {Antoniuk}, \citenamefont {Lettner}, \citenamefont {Ovvyan},
  \citenamefont {Waltrich}, \citenamefont {Gruhler}, \citenamefont {Davydov},
  \citenamefont {Agafonov}, \citenamefont {Pernice},\ and\ \citenamefont
  {Kubanek}}]{fehler_hybrid_2021}%
  \BibitemOpen
  \bibfield  {author} {\bibinfo {author} {\bibfnamefont {K.~G.}\ \bibnamefont
  {Fehler}}, \bibinfo {author} {\bibfnamefont {L.}~\bibnamefont {Antoniuk}},
  \bibinfo {author} {\bibfnamefont {N.}~\bibnamefont {Lettner}}, \bibinfo
  {author} {\bibfnamefont {A.~P.}\ \bibnamefont {Ovvyan}}, \bibinfo {author}
  {\bibfnamefont {R.}~\bibnamefont {Waltrich}}, \bibinfo {author}
  {\bibfnamefont {N.}~\bibnamefont {Gruhler}}, \bibinfo {author} {\bibfnamefont
  {V.~A.}\ \bibnamefont {Davydov}}, \bibinfo {author} {\bibfnamefont {V.~N.}\
  \bibnamefont {Agafonov}}, \bibinfo {author} {\bibfnamefont {W.~H.~P.}\
  \bibnamefont {Pernice}},\ and\ \bibinfo {author} {\bibfnamefont
  {A.}~\bibnamefont {Kubanek}},\ }\bibfield  {title} {\bibinfo {title} {Hybrid
  {{Quantum Photonics Based}} on {{Artificial Atoms Placed Inside One Hole}} of
  a {{Photonic Crystal Cavity}}},\ }\href
  {https://doi.org/10.1021/acsphotonics.1c00530} {\bibfield  {journal}
  {\bibinfo  {journal} {ACS Photonics}\ }\textbf {\bibinfo {volume} {8}},\
  \bibinfo {pages} {2635} (\bibinfo {year} {2021})}\BibitemShut {NoStop}%
\bibitem [{\citenamefont {H{\"a}u{\ss}ler}\ \emph
  {et~al.}(2019{\natexlab{b}})\citenamefont {H{\"a}u{\ss}ler}, \citenamefont
  {Hartung}, \citenamefont {Fehler}, \citenamefont {Antoniuk}, \citenamefont
  {Kulikova}, \citenamefont {Davydov}, \citenamefont {Agafonov}, \citenamefont
  {Jelezko},\ and\ \citenamefont {Kubanek}}]{hausler_preparing_2019}%
  \BibitemOpen
  \bibfield  {author} {\bibinfo {author} {\bibfnamefont {S.}~\bibnamefont
  {H{\"a}u{\ss}ler}}, \bibinfo {author} {\bibfnamefont {L.}~\bibnamefont
  {Hartung}}, \bibinfo {author} {\bibfnamefont {K.~G.}\ \bibnamefont {Fehler}},
  \bibinfo {author} {\bibfnamefont {L.}~\bibnamefont {Antoniuk}}, \bibinfo
  {author} {\bibfnamefont {L.~F.}\ \bibnamefont {Kulikova}}, \bibinfo {author}
  {\bibfnamefont {V.~A.}\ \bibnamefont {Davydov}}, \bibinfo {author}
  {\bibfnamefont {V.~N.}\ \bibnamefont {Agafonov}}, \bibinfo {author}
  {\bibfnamefont {F.}~\bibnamefont {Jelezko}},\ and\ \bibinfo {author}
  {\bibfnamefont {A.}~\bibnamefont {Kubanek}},\ }\bibfield  {title} {\bibinfo
  {title} {Preparing single {{SiV}}- center in nanodiamonds for external,
  optical coupling with access to all degrees of freedom},\ }\href
  {https://doi.org/10.1088/1367-2630/ab4cf7} {\bibfield  {journal} {\bibinfo
  {journal} {New Journal of Physics}\ }\textbf {\bibinfo {volume} {21}},\
  \bibinfo {pages} {103047} (\bibinfo {year} {2019}{\natexlab{b}})}\BibitemShut
  {NoStop}%
\bibitem [{\citenamefont {Dolan}\ \emph {et~al.}(2010)\citenamefont {Dolan},
  \citenamefont {Hughes}, \citenamefont {Grazioso}, \citenamefont {Patton},\
  and\ \citenamefont {Smith}}]{dolan_femtoliter_2010}%
  \BibitemOpen
  \bibfield  {author} {\bibinfo {author} {\bibfnamefont {P.~R.}\ \bibnamefont
  {Dolan}}, \bibinfo {author} {\bibfnamefont {G.~M.}\ \bibnamefont {Hughes}},
  \bibinfo {author} {\bibfnamefont {F.}~\bibnamefont {Grazioso}}, \bibinfo
  {author} {\bibfnamefont {B.~R.}\ \bibnamefont {Patton}},\ and\ \bibinfo
  {author} {\bibfnamefont {J.~M.}\ \bibnamefont {Smith}},\ }\bibfield  {title}
  {\bibinfo {title} {Femtoliter tunable optical cavity arrays},\ }\href
  {https://doi.org/10.1364/OL.35.003556} {\bibfield  {journal} {\bibinfo
  {journal} {Optics Letters}\ }\textbf {\bibinfo {volume} {35}},\ \bibinfo
  {pages} {3556} (\bibinfo {year} {2010})}\BibitemShut {NoStop}%
\bibitem [{\citenamefont {Kelkar}\ \emph {et~al.}(2015)\citenamefont {Kelkar},
  \citenamefont {Wang}, \citenamefont {{Mart{\'i}n-Cano}}, \citenamefont
  {Hoffmann}, \citenamefont {Christiansen}, \citenamefont {G{\"o}tzinger},\
  and\ \citenamefont {Sandoghdar}}]{kelkar_sensing_2015-1}%
  \BibitemOpen
  \bibfield  {author} {\bibinfo {author} {\bibfnamefont {H.}~\bibnamefont
  {Kelkar}}, \bibinfo {author} {\bibfnamefont {D.}~\bibnamefont {Wang}},
  \bibinfo {author} {\bibfnamefont {D.}~\bibnamefont {{Mart{\'i}n-Cano}}},
  \bibinfo {author} {\bibfnamefont {B.}~\bibnamefont {Hoffmann}}, \bibinfo
  {author} {\bibfnamefont {S.}~\bibnamefont {Christiansen}}, \bibinfo {author}
  {\bibfnamefont {S.}~\bibnamefont {G{\"o}tzinger}},\ and\ \bibinfo {author}
  {\bibfnamefont {V.}~\bibnamefont {Sandoghdar}},\ }\bibfield  {title}
  {\bibinfo {title} {Sensing {{Nanoparticles}} with a {{Cantilever-Based
  Scannable Optical Cavity}} of {{Low Finesse}} and {{Sub-}} {$\lambda$} 3
  {{Volume}}},\ }\href {https://doi.org/10.1103/PhysRevApplied.4.054010}
  {\bibfield  {journal} {\bibinfo  {journal} {Physical Review Applied}\
  }\textbf {\bibinfo {volume} {4}},\ \bibinfo {pages} {054010} (\bibinfo {year}
  {2015})}\BibitemShut {NoStop}%
\bibitem [{\citenamefont {Dolan}\ \emph {et~al.}(2018)\citenamefont {Dolan},
  \citenamefont {Adekanye}, \citenamefont {Trichet}, \citenamefont {Johnson},
  \citenamefont {Flatten}, \citenamefont {Chen}, \citenamefont {Weng},
  \citenamefont {Hunger}, \citenamefont {Chang}, \citenamefont {Castelletto},\
  and\ \citenamefont {Smith.}}]{dolan_robust_2018}%
  \BibitemOpen
  \bibfield  {author} {\bibinfo {author} {\bibfnamefont {P.~R.}\ \bibnamefont
  {Dolan}}, \bibinfo {author} {\bibfnamefont {S.}~\bibnamefont {Adekanye}},
  \bibinfo {author} {\bibfnamefont {A.~A.~P.}\ \bibnamefont {Trichet}},
  \bibinfo {author} {\bibfnamefont {S.}~\bibnamefont {Johnson}}, \bibinfo
  {author} {\bibfnamefont {L.~C.}\ \bibnamefont {Flatten}}, \bibinfo {author}
  {\bibfnamefont {Y.~C.}\ \bibnamefont {Chen}}, \bibinfo {author}
  {\bibfnamefont {L.}~\bibnamefont {Weng}}, \bibinfo {author} {\bibfnamefont
  {D.}~\bibnamefont {Hunger}}, \bibinfo {author} {\bibfnamefont {H.-C.}\
  \bibnamefont {Chang}}, \bibinfo {author} {\bibfnamefont {S.}~\bibnamefont
  {Castelletto}},\ and\ \bibinfo {author} {\bibfnamefont {J.~M.}\ \bibnamefont
  {Smith.}},\ }\bibfield  {title} {\bibinfo {title} {Robust, tunable, and high
  purity triggered single photon source at room temperature using a
  nitrogen-vacancy defect in diamond in an open microcavity},\ }\href
  {https://doi.org/10.1364/OE.26.007056} {\bibfield  {journal} {\bibinfo
  {journal} {Optics Express}\ }\textbf {\bibinfo {volume} {26}},\ \bibinfo
  {pages} {7056} (\bibinfo {year} {2018})}\BibitemShut {NoStop}%
\bibitem [{\citenamefont {Neu}\ \emph {et~al.}(2011)\citenamefont {Neu},
  \citenamefont {Steinmetz}, \citenamefont {{Riedrich-M{\"o}ller}},
  \citenamefont {Gsell}, \citenamefont {Fischer}, \citenamefont {Schreck},\
  and\ \citenamefont {Becher}}]{neu_single_2011}%
  \BibitemOpen
  \bibfield  {author} {\bibinfo {author} {\bibfnamefont {E.}~\bibnamefont
  {Neu}}, \bibinfo {author} {\bibfnamefont {D.}~\bibnamefont {Steinmetz}},
  \bibinfo {author} {\bibfnamefont {J.}~\bibnamefont {{Riedrich-M{\"o}ller}}},
  \bibinfo {author} {\bibfnamefont {S.}~\bibnamefont {Gsell}}, \bibinfo
  {author} {\bibfnamefont {M.}~\bibnamefont {Fischer}}, \bibinfo {author}
  {\bibfnamefont {M.}~\bibnamefont {Schreck}},\ and\ \bibinfo {author}
  {\bibfnamefont {C.}~\bibnamefont {Becher}},\ }\bibfield  {title} {\bibinfo
  {title} {Single photon emission from silicon-vacancy colour centres in
  chemical vapour deposition nano-diamonds on iridium},\ }\href
  {https://doi.org/10.1088/1367-2630/13/2/025012} {\bibfield  {journal}
  {\bibinfo  {journal} {New Journal of Physics}\ }\textbf {\bibinfo {volume}
  {13}},\ \bibinfo {pages} {025012} (\bibinfo {year} {2011})}\BibitemShut
  {NoStop}%
\bibitem [{\citenamefont {Zhang}\ \emph {et~al.}(2018)\citenamefont {Zhang},
  \citenamefont {Sun}, \citenamefont {Burek}, \citenamefont {Dory},
  \citenamefont {Tzeng}, \citenamefont {Fischer}, \citenamefont {Kelaita},
  \citenamefont {Lagoudakis}, \citenamefont {Radulaski}, \citenamefont {Shen},
  \citenamefont {Melosh}, \citenamefont {Chu}, \citenamefont {Lon{\v c}ar},\
  and\ \citenamefont {Vu{\v c}kovi{\'c}}}]{zhang_strongly_2018}%
  \BibitemOpen
  \bibfield  {author} {\bibinfo {author} {\bibfnamefont {J.~L.}\ \bibnamefont
  {Zhang}}, \bibinfo {author} {\bibfnamefont {S.}~\bibnamefont {Sun}}, \bibinfo
  {author} {\bibfnamefont {M.~J.}\ \bibnamefont {Burek}}, \bibinfo {author}
  {\bibfnamefont {C.}~\bibnamefont {Dory}}, \bibinfo {author} {\bibfnamefont
  {Y.-K.}\ \bibnamefont {Tzeng}}, \bibinfo {author} {\bibfnamefont {K.~A.}\
  \bibnamefont {Fischer}}, \bibinfo {author} {\bibfnamefont {Y.}~\bibnamefont
  {Kelaita}}, \bibinfo {author} {\bibfnamefont {K.~G.}\ \bibnamefont
  {Lagoudakis}}, \bibinfo {author} {\bibfnamefont {M.}~\bibnamefont
  {Radulaski}}, \bibinfo {author} {\bibfnamefont {Z.-X.}\ \bibnamefont {Shen}},
  \bibinfo {author} {\bibfnamefont {N.~A.}\ \bibnamefont {Melosh}}, \bibinfo
  {author} {\bibfnamefont {S.}~\bibnamefont {Chu}}, \bibinfo {author}
  {\bibfnamefont {M.}~\bibnamefont {Lon{\v c}ar}},\ and\ \bibinfo {author}
  {\bibfnamefont {J.}~\bibnamefont {Vu{\v c}kovi{\'c}}},\ }\bibfield  {title}
  {\bibinfo {title} {Strongly {{Cavity-Enhanced Spontaneous Emission}} from
  {{Silicon-Vacancy Centers}} in {{Diamond}}},\ }\href
  {https://doi.org/10.1021/acs.nanolett.7b05075} {\bibfield  {journal}
  {\bibinfo  {journal} {Nano Letters}\ }\textbf {\bibinfo {volume} {18}},\
  \bibinfo {pages} {1360} (\bibinfo {year} {2018})}\BibitemShut {NoStop}%
\bibitem [{\citenamefont {Meesala}\ \emph {et~al.}(2018)\citenamefont
  {Meesala}, \citenamefont {Sohn}, \citenamefont {Pingault}, \citenamefont
  {Shao}, \citenamefont {Atikian}, \citenamefont {Holzgrafe}, \citenamefont
  {G{\"u}ndo{\u g}an}, \citenamefont {Stavrakas}, \citenamefont {Sipahigil},
  \citenamefont {Chia}, \citenamefont {Evans}, \citenamefont {Burek},
  \citenamefont {Zhang}, \citenamefont {Wu}, \citenamefont {Pacheco},
  \citenamefont {Abraham}, \citenamefont {Bielejec}, \citenamefont {Lukin},
  \citenamefont {Atat{\"u}re},\ and\ \citenamefont {Lon{\v
  c}ar}}]{meesala_strain_2018}%
  \BibitemOpen
  \bibfield  {author} {\bibinfo {author} {\bibfnamefont {S.}~\bibnamefont
  {Meesala}}, \bibinfo {author} {\bibfnamefont {Y.-I.}\ \bibnamefont {Sohn}},
  \bibinfo {author} {\bibfnamefont {B.}~\bibnamefont {Pingault}}, \bibinfo
  {author} {\bibfnamefont {L.}~\bibnamefont {Shao}}, \bibinfo {author}
  {\bibfnamefont {H.~A.}\ \bibnamefont {Atikian}}, \bibinfo {author}
  {\bibfnamefont {J.}~\bibnamefont {Holzgrafe}}, \bibinfo {author}
  {\bibfnamefont {M.}~\bibnamefont {G{\"u}ndo{\u g}an}}, \bibinfo {author}
  {\bibfnamefont {C.}~\bibnamefont {Stavrakas}}, \bibinfo {author}
  {\bibfnamefont {A.}~\bibnamefont {Sipahigil}}, \bibinfo {author}
  {\bibfnamefont {C.}~\bibnamefont {Chia}}, \bibinfo {author} {\bibfnamefont
  {R.}~\bibnamefont {Evans}}, \bibinfo {author} {\bibfnamefont {M.~J.}\
  \bibnamefont {Burek}}, \bibinfo {author} {\bibfnamefont {M.}~\bibnamefont
  {Zhang}}, \bibinfo {author} {\bibfnamefont {L.}~\bibnamefont {Wu}}, \bibinfo
  {author} {\bibfnamefont {J.~L.}\ \bibnamefont {Pacheco}}, \bibinfo {author}
  {\bibfnamefont {J.}~\bibnamefont {Abraham}}, \bibinfo {author} {\bibfnamefont
  {E.}~\bibnamefont {Bielejec}}, \bibinfo {author} {\bibfnamefont {M.~D.}\
  \bibnamefont {Lukin}}, \bibinfo {author} {\bibfnamefont {M.}~\bibnamefont
  {Atat{\"u}re}},\ and\ \bibinfo {author} {\bibfnamefont {M.}~\bibnamefont
  {Lon{\v c}ar}},\ }\bibfield  {title} {\bibinfo {title} {Strain engineering of
  the silicon-vacancy center in diamond},\ }\href
  {https://doi.org/10.1103/PhysRevB.97.205444} {\bibfield  {journal} {\bibinfo
  {journal} {Physical Review B}\ }\textbf {\bibinfo {volume} {97}},\ \bibinfo
  {pages} {205444} (\bibinfo {year} {2018})}\BibitemShut {NoStop}%
\bibitem [{\citenamefont {Rogers}\ \emph
  {et~al.}(2014{\natexlab{b}})\citenamefont {Rogers}, \citenamefont {Jahnke},
  \citenamefont {Metsch}, \citenamefont {Sipahigil}, \citenamefont {Binder},
  \citenamefont {Teraji}, \citenamefont {Sumiya}, \citenamefont {Isoya},
  \citenamefont {Lukin}, \citenamefont {Hemmer},\ and\ \citenamefont
  {Jelezko}}]{rogers_all-optical_2014}%
  \BibitemOpen
  \bibfield  {author} {\bibinfo {author} {\bibfnamefont {L.~J.}\ \bibnamefont
  {Rogers}}, \bibinfo {author} {\bibfnamefont {K.~D.}\ \bibnamefont {Jahnke}},
  \bibinfo {author} {\bibfnamefont {M.~H.}\ \bibnamefont {Metsch}}, \bibinfo
  {author} {\bibfnamefont {A.}~\bibnamefont {Sipahigil}}, \bibinfo {author}
  {\bibfnamefont {J.~M.}\ \bibnamefont {Binder}}, \bibinfo {author}
  {\bibfnamefont {T.}~\bibnamefont {Teraji}}, \bibinfo {author} {\bibfnamefont
  {H.}~\bibnamefont {Sumiya}}, \bibinfo {author} {\bibfnamefont
  {J.}~\bibnamefont {Isoya}}, \bibinfo {author} {\bibfnamefont {M.~D.}\
  \bibnamefont {Lukin}}, \bibinfo {author} {\bibfnamefont {P.}~\bibnamefont
  {Hemmer}},\ and\ \bibinfo {author} {\bibfnamefont {F.}~\bibnamefont
  {Jelezko}},\ }\bibfield  {title} {\bibinfo {title} {All-{{Optical
  Initialization}}, {{Readout}}, and {{Coherent Preparation}} of {{Single
  Silicon-Vacancy Spins}} in {{Diamond}}},\ }\href
  {https://doi.org/10.1103/PhysRevLett.113.263602} {\bibfield  {journal}
  {\bibinfo  {journal} {Physical Review Letters}\ }\textbf {\bibinfo {volume}
  {113}},\ \bibinfo {pages} {263602} (\bibinfo {year}
  {2014}{\natexlab{b}})}\BibitemShut {NoStop}%
\bibitem [{\citenamefont {Rondin}\ \emph {et~al.}(2014)\citenamefont {Rondin},
  \citenamefont {Tetienne}, \citenamefont {Hingant}, \citenamefont {Roch},
  \citenamefont {Maletinsky},\ and\ \citenamefont
  {Jacques}}]{rondin_magnetometry_2014}%
  \BibitemOpen
  \bibfield  {author} {\bibinfo {author} {\bibfnamefont {L.}~\bibnamefont
  {Rondin}}, \bibinfo {author} {\bibfnamefont {J.-P.}\ \bibnamefont
  {Tetienne}}, \bibinfo {author} {\bibfnamefont {T.}~\bibnamefont {Hingant}},
  \bibinfo {author} {\bibfnamefont {J.-F.}\ \bibnamefont {Roch}}, \bibinfo
  {author} {\bibfnamefont {P.}~\bibnamefont {Maletinsky}},\ and\ \bibinfo
  {author} {\bibfnamefont {V.}~\bibnamefont {Jacques}},\ }\bibfield  {title}
  {\bibinfo {title} {Magnetometry with nitrogen-vacancy defects in diamond},\
  }\href {https://doi.org/10.1088/0034-4885/77/5/056503} {\bibfield  {journal}
  {\bibinfo  {journal} {Reports on Progress in Physics}\ }\textbf {\bibinfo
  {volume} {77}},\ \bibinfo {pages} {056503} (\bibinfo {year}
  {2014})}\BibitemShut {NoStop}%
\bibitem [{\citenamefont {Stepanov}\ \emph {et~al.}(2015)\citenamefont
  {Stepanov}, \citenamefont {Cho}, \citenamefont {Abeywardana},\ and\
  \citenamefont {Takahashi}}]{stepanov_high-frequency_2015}%
  \BibitemOpen
  \bibfield  {author} {\bibinfo {author} {\bibfnamefont {V.}~\bibnamefont
  {Stepanov}}, \bibinfo {author} {\bibfnamefont {F.~H.}\ \bibnamefont {Cho}},
  \bibinfo {author} {\bibfnamefont {C.}~\bibnamefont {Abeywardana}},\ and\
  \bibinfo {author} {\bibfnamefont {S.}~\bibnamefont {Takahashi}},\ }\bibfield
  {title} {\bibinfo {title} {High-frequency and high-field optically detected
  magnetic resonance of nitrogen-vacancy centers in diamond},\ }\href
  {https://doi.org/10.1063/1.4908528} {\bibfield  {journal} {\bibinfo
  {journal} {Applied Physics Letters}\ }\textbf {\bibinfo {volume} {106}},\
  \bibinfo {pages} {063111} (\bibinfo {year} {2015})}\BibitemShut {NoStop}%
\bibitem [{\citenamefont {{van Loock}}\ \emph {et~al.}(2020)\citenamefont {{van
  Loock}}, \citenamefont {Alt}, \citenamefont {Becher}, \citenamefont {Benson},
  \citenamefont {Boche}, \citenamefont {Deppe}, \citenamefont {Eschner},
  \citenamefont {H{\"o}fling}, \citenamefont {Meschede}, \citenamefont
  {Michler}, \citenamefont {Schmidt},\ and\ \citenamefont
  {Weinfurter}}]{van_loock_extending_2020}%
  \BibitemOpen
  \bibfield  {author} {\bibinfo {author} {\bibfnamefont {P.}~\bibnamefont {{van
  Loock}}}, \bibinfo {author} {\bibfnamefont {W.}~\bibnamefont {Alt}}, \bibinfo
  {author} {\bibfnamefont {C.}~\bibnamefont {Becher}}, \bibinfo {author}
  {\bibfnamefont {O.}~\bibnamefont {Benson}}, \bibinfo {author} {\bibfnamefont
  {H.}~\bibnamefont {Boche}}, \bibinfo {author} {\bibfnamefont
  {C.}~\bibnamefont {Deppe}}, \bibinfo {author} {\bibfnamefont
  {J.}~\bibnamefont {Eschner}}, \bibinfo {author} {\bibfnamefont
  {S.}~\bibnamefont {H{\"o}fling}}, \bibinfo {author} {\bibfnamefont
  {D.}~\bibnamefont {Meschede}}, \bibinfo {author} {\bibfnamefont
  {P.}~\bibnamefont {Michler}}, \bibinfo {author} {\bibfnamefont
  {F.}~\bibnamefont {Schmidt}},\ and\ \bibinfo {author} {\bibfnamefont
  {H.}~\bibnamefont {Weinfurter}},\ }\bibfield  {title} {\bibinfo {title}
  {Extending {{Quantum Links}}: {{Modules}} for {{Fiber-}} and {{Memory-Based
  Quantum Repeaters}}},\ }\href {https://doi.org/10.1002/qute.201900141}
  {\bibfield  {journal} {\bibinfo  {journal} {Advanced Quantum Technologies}\
  }\textbf {\bibinfo {volume} {3}},\ \bibinfo {pages} {1900141} (\bibinfo
  {year} {2020})}\BibitemShut {NoStop}%
\bibitem [{\citenamefont {Lee}\ \emph {et~al.}(2022)\citenamefont {Lee},
  \citenamefont {Bersin}, \citenamefont {Dahlberg}, \citenamefont {Wehner},\
  and\ \citenamefont {Englund}}]{lee_quantum_2022}%
  \BibitemOpen
  \bibfield  {author} {\bibinfo {author} {\bibfnamefont {Y.}~\bibnamefont
  {Lee}}, \bibinfo {author} {\bibfnamefont {E.}~\bibnamefont {Bersin}},
  \bibinfo {author} {\bibfnamefont {A.}~\bibnamefont {Dahlberg}}, \bibinfo
  {author} {\bibfnamefont {S.}~\bibnamefont {Wehner}},\ and\ \bibinfo {author}
  {\bibfnamefont {D.}~\bibnamefont {Englund}},\ }\bibfield  {title} {\bibinfo
  {title} {A quantum router architecture for high-fidelity entanglement flows
  in quantum networks},\ }\href {https://doi.org/10.1038/s41534-022-00582-8}
  {\bibfield  {journal} {\bibinfo  {journal} {npj Quantum Information}\
  }\textbf {\bibinfo {volume} {8}},\ \bibinfo {pages} {1} (\bibinfo {year}
  {2022})}\BibitemShut {NoStop}%
\bibitem [{\citenamefont {Binder}\ \emph {et~al.}(2017)\citenamefont {Binder},
  \citenamefont {Stark}, \citenamefont {Tomek}, \citenamefont {Scheuer},
  \citenamefont {Frank}, \citenamefont {Jahnke}, \citenamefont {M{\"u}ller},
  \citenamefont {Schmitt}, \citenamefont {Metsch}, \citenamefont {Unden},
  \citenamefont {Gehring}, \citenamefont {Huck}, \citenamefont {Andersen},
  \citenamefont {Rogers},\ and\ \citenamefont {Jelezko}}]{binder_qudi_2017}%
  \BibitemOpen
  \bibfield  {author} {\bibinfo {author} {\bibfnamefont {J.~M.}\ \bibnamefont
  {Binder}}, \bibinfo {author} {\bibfnamefont {A.}~\bibnamefont {Stark}},
  \bibinfo {author} {\bibfnamefont {N.}~\bibnamefont {Tomek}}, \bibinfo
  {author} {\bibfnamefont {J.}~\bibnamefont {Scheuer}}, \bibinfo {author}
  {\bibfnamefont {F.}~\bibnamefont {Frank}}, \bibinfo {author} {\bibfnamefont
  {K.~D.}\ \bibnamefont {Jahnke}}, \bibinfo {author} {\bibfnamefont
  {C.}~\bibnamefont {M{\"u}ller}}, \bibinfo {author} {\bibfnamefont
  {S.}~\bibnamefont {Schmitt}}, \bibinfo {author} {\bibfnamefont {M.~H.}\
  \bibnamefont {Metsch}}, \bibinfo {author} {\bibfnamefont {T.}~\bibnamefont
  {Unden}}, \bibinfo {author} {\bibfnamefont {T.}~\bibnamefont {Gehring}},
  \bibinfo {author} {\bibfnamefont {A.}~\bibnamefont {Huck}}, \bibinfo {author}
  {\bibfnamefont {U.~L.}\ \bibnamefont {Andersen}}, \bibinfo {author}
  {\bibfnamefont {L.~J.}\ \bibnamefont {Rogers}},\ and\ \bibinfo {author}
  {\bibfnamefont {F.}~\bibnamefont {Jelezko}},\ }\bibfield  {title} {\bibinfo
  {title} {Qudi: {{A}} modular python suite for experiment control and data
  processing},\ }\href {https://doi.org/10.1016/j.softx.2017.02.001} {\bibfield
   {journal} {\bibinfo  {journal} {SoftwareX}\ }\textbf {\bibinfo {volume}
  {6}},\ \bibinfo {pages} {85} (\bibinfo {year} {2017})}\BibitemShut {NoStop}%
\end{thebibliography}%

\end{document}